\documentclass[12pt]{iopart}
\usepackage{graphicx}
\expandafter\let\csname equation*\endcsname\relax
\expandafter\let\csname endequation*\endcsname\relax
\usepackage{amsmath}
\usepackage{color,soul}
\usepackage{algorithm}
\usepackage{algpseudocode}
\usepackage{bm}
\usepackage{esvect}
\usepackage{accents}
\usepackage{cite}
\newcommand*{\dt}[1]{%
  \accentset{\mbox{\large\bfseries .}}{#1}}
\newcommand*{\ddt}[1]{%
  \accentset{\mbox{\large\bfseries .\hspace{-0.25ex}.}}{#1}}

\begin{document}

\title[MICROSCOPE data analysis]{MICROSCOPE mission: Data analysis principle}

\author{Joel Berg\'e$^1$, Quentin Baghi$^{1,2}$\footnote{Current address: NASA Goddard Space Flight Center, Greenbelt, MD 20771, USA}, Emilie Hardy$^1$, Gilles M\'etris$^2$, Alain Robert$^3$, Manuel Rodrigues$^1$, Pierre Touboul$^1$, Ratana Chhun$^1$, Pierre-Yves Guidotti$^3$\footnote{Current address: Airbus Defence and Space, 31 rue des Cosmonautes, 31402 Toulouse, France}, Sandrine Pires$^4$, Serge Reynaud$^{5}$, Laura Serron$^2$, Jean-Michel Travert$^6$}

\address{$^1$ DPHY, ONERA, Universit\'e Paris Saclay, F-92322 Ch\^atillon, France}
\address{$^2$ Universit\'e C\^ote d'Azur, Observatoire de la C\^ote d'Azur, CNRS, IRD, G\'eoazur, 250 avenue Albert Einstein, F-06560 Valbonne, France}
\address{$^3$ CNES Toulouse, 18 avenue Edouard Belin - 31401 Toulouse Cedex 9, France}
\address{$^4$ Laboratoire AIM, CEA/DSM--CNRS, Universit\'e Paris Diderot, IRFU/SAp, CEA Saclay,  Orme des Merisiers,  91191 Gif-sur-Yvette, France}
\address{$^{5}$ Laboratoire Kastler Brossel, UPMC-Sorbonne Universit\'e, CNRS, ENS-PSL University, Coll\`ege de France, 75252 Paris, France}
\address{$^6$ Altran technologies, 17 Avenue Didier Daurat, 31700 Blagnac, France}
\ead{joel.berge@onera.fr}
\vspace{10pt}
\begin{indented}
\item[]December 2020
\end{indented}

\begin{abstract}
After performing highly sensitive acceleration measurements during two years of drag-free flight around the Earth, MICROSCOPE provided the best constraint on the Weak Equivalence Principle (WEP) to date. Beside being a technological challenge, this experiment required a specialised data analysis pipeline to look for a potential small signal burried in the noise, possibly plagued by instrumental defects, missing data and glitches. This paper describes the frequency-domain iterative least-square technique that we developed for MICROSCOPE. In particular, using numerical simulations, we prove that our estimator is unbiased and provides correct error bars. This paper therefore justifies the robustness of the WEP measurements given by MICROSCOPE.
\end{abstract}

%

\submitto{\CQG}
%
%
%

\section{Introduction} \label{sect_intro}

MICROSCOPE's test of the Weak Equivalence Principle (WEP) is based on the comparison of the acceleration of two concentric cylindrical test masses of different composition as they orbit the Earth \cite{touboul17,touboul19,touboulcqg0}.
Thus, the gravity that pulls the test masses being sourced by the Earth, any Equivalence Principle violation (EPV) signal will be proportional to the Earth gravity acceleration. 
More precisely, as it is measured along the main axis of the cylindrical masses, the signal we look for can be expected to be proportional to the Earth's gravity acceleration modulated by the motion and the attitude of the satellite around the Earth, resulting in a periodic signal with a well-known frequency (noted $f_{\rm EP}$ in this paper).

The main MICROSCOPE data consists in time series of accelerations measured by two concentric accelerometers \cite{rodriguescqg1}. To look for an EPV, it is then enough to look for a non-zero signal at the $f_{\rm EP}$ frequency in the difference of those two time series. Obviously, this process is impacted by instrumental noise and systematics (either instrumental or environmental). Thus, it consists in correcting the measured time series from --calibrated and/or modelled-- instrumental and environmental systematics, before seeking a possible periodic signal amounting to a violation of the WEP in coloured-noise-dominated data.

As described in Ref. \cite{rodriguescqg4}, the MICROSCOPE mission is divided in different measurement sessions. Sessions represent a time span during which the satellite and the instrument keep the same configuration (spin, drag-free control law...). Some of these sessions are directly devoted to the WEP test (called ``EP sessions'' in this paper) while others (``calibration sessions'') are used to calibrate or characterise the instrument. EP sessions are the longest, most of them lasting 120 orbital periods (about 8 days), while calibration sessions typically last a few orbits.

In this paper, we present MICROSCOPE's data analysis pipeline. After recalling the measurement equation in Sect. \ref{sect_measurement}, we present our   frequency-domain iterative ordinary least squares (OLS) algorithm and its mathematical background in Sect. \ref{sect_pestimation}. We then use ``worst-case'' simulations from a hybrid software-hardware MICROSCOPE simulator to discuss the optimal way to correct for gaps in MICROSCOPE data, in Sect. \ref{sect_transients}. Finally, in Sect. \ref{sect_pysimula}, we show with well-controlled numerical simulations that our pipeline provides unbiased estimates and reliable error bars in the presence of instrumental systematics. Combined with previous works where we showed how we can successfully deal with missing data, those results prove that our data analysis pipeline allows us to reliably measure an EPV, if any. 
We conclude in Sect. \ref{sect_conclusion}. A detailed appendix discusses uncertainty propagation.
Note that this paper is only about the MICROSCOPE data analysis methodology, and does not present real data. For the real data analysis (using the methods described in this paper), see Refs \cite{touboulcqg0, hardycqg6, bergecqg8, metriscqg9}.

The terminology and notations used in this paper, as well as all observables, are defined in Ref. \cite{rodriguescqg1}. In particular, we define the common-mode (resp. differential-mode) of a given observable or parameter as the half-sum (resp. half-difference) of this observable/parameter for both test masses, $o^{(c,d)} = (o^{(1)}\pm{}o^{(2)})/2$, and we use the convention $\mathbf{o}$ to denote a vector and $[\mathbf o]$ to denote a second-order tensor, respectively. We note the time derivative of observable $o$ as $\dt{o}$.

\section{Measurement principle} \label{sect_measurement}

\subsection{Measurement equation}

MICROSCOPE looks for an EPV by monitoring the difference in accelerations potentially undergone by the two test masses of a differential accelerometer \cite{touboul17,touboul19,touboulcqg0,rodriguescqg1}.
In an ideal case, the measurement equation is then straightforward to establish, since the difference of acceleration of the two test masses is expected to be proportional to Earth gravity field ${\mathbf g}$, ${\mathbf \Gamma^{(d)}} \equiv {\mathbf \Gamma^{(1)}} - {\mathbf \Gamma^{(2)}} = \delta {\mathbf g}$, where $\delta$ is the (approximate) E\"otv\"os parameter that we wish to estimate. 
However, the instrument is not perfect: for instance, scale factors are not exactly unit and test masses are imperfectly centered and aligned with respect to each other and to the satellite's frame of reference, such that the common-mode and differential-mode sensitivity matrices $[\mathbf{a_c}]$ and $[\mathbf{a_d}]$ are not the identity and null matrices, respectively.
Furthermore, since the satellite rotates it imparts a Coriolis acceleration on the test masses. 

The measurement equation is therefore much more complicated than in the ideal case. Ref. \cite{rodriguescqg1} establishes it in full generality, introducing and taking into account all instrumental defects and their notation. We measure the differential acceleration along the test masses' sensitive axis ($x$-coordinate in the instrument's frame, see Fig. 1 of Ref. \cite{liorzoucqg2}), such that
\begin{multline}
  \label{eq_xacc}
      \Gamma_x^{(d)} = 2 \tilde{b}_x^{(d)}+ a_{c11}{ \delta}{ g_x}+ a_{c12}{ \delta} {g_y}+a_{c13}{ \delta} { g_z}+{\Delta'_{x} }{{ S_{xx}}}+{\Delta'_{y} }{{ S_{xy}}}+{\Delta'_{z} }{{ S_{xz}}}\\
+{\left( a_{c13}\Delta'_{y}+a_{c12} \Delta'_{z} \right) }{{ S_{yz}}}+{a_{c12} \Delta'_{y}}{{ S_{yy}}}+{a_{c13} \Delta'_{z}}{{ S_{zz}}}\\
+\left({-a_{c13}\Delta'_{y}+a_{c12}  \Delta'_{z} }+2c_{d11}\right){\dt\Omega_x}-\left({\Delta'_{z}-2 a_{c13} \Delta'_{x}}-2 c_{d12} \right){\dt\Omega_y}\\
+\left({\Delta'_{y}-2 a_{c12} \Delta'_{x}}+2 c_{d13} \right){\dt\Omega_z}+2\left( { a_{d11}}{ \tilde\Gamma^{(c)}_x}+{ a_{d12}}{
    \tilde\Gamma^{(c)}_y}+{ a_{d13}}{ \tilde\Gamma^{(c)}_z}\right) \\
+ 2 \dt{\Delta}'_x \Omega_x - 2 \dt{\Delta}'_z \Omega_y + 2 \dt{\Delta}'_y \Omega_z -a_{c11} \ddt{\Delta}_x - a_{c12} \ddt{\Delta}_y - a_{c13} \ddt{\Delta}_z \\
+{ \tilde K^{(1)}_{2xx}}\left( {\tilde\Gamma^{(1)}_x}\right)^2-{ \tilde K^{(2)}_{2xx}}\left( {\tilde\Gamma^{(2)}_x}\right)^2
+ 2 n_x^{(d)},
\end{multline}
where $\left[{\mathbf T}\right]$ the Earth gravity gradient tensor (GGT) in the instrument's frame, $\left[{\mathbf {In}}\right] = \left[\dt{\boldsymbol\Omega} \right] + \left[\boldsymbol\Omega \right]\left[\boldsymbol\Omega \right]$ the gradient of inertia tensor, $\left[{\mathbf S}\right]$ the symmetric part of the $\left[{\mathbf T}\right]-\left[{\mathbf {In}}\right]$ matrix, $\left[\boldsymbol\Omega \right]$ the angular velocity tensor of the satellite, and $\mathbf \Delta$ the vector between the center of the two test masses (called ``offcentering vector'' hereafter). 

In Eq. (\ref{eq_xacc}), $\tilde{ \mathbf\Gamma}^{(c)} =  \mathbf \Gamma^{(c)} - \mathbf n^{(c)}$ is the noise-free common-mode measured acceleration, $\left[a_{c11}, a_{c12}, a_{c13} \right]$ is the first row of the common-mode sensitivity matrix, $\left[a_{d11}, a_{d12}, a_{d13} \right]$ is the first row of the differential-mode sensitivity matrix, $\left[c_{d11}, c_{d12}, c_{d13} \right]$ is the first row of the differential-mode sensitivity matrix to the angular acceleration, $\tilde{b}_x^{(d)} = b_{0x}^{d} + a_{c11}b_{1x}^{(d)} + a_{c12}b_{1y}^{(d)} + a_{c13}b_{1z}^{(d)}$ (with $\mathbf{b_0}^{(d)}$ the differential electrostatic bias and $\mathbf {b_1}^{(d)}$ the difference of mechanical perturbations acting on the two test masses). For convenience, in the remainder of this paper, we denote $\delta_x\equiv a_{c11} \delta$,  $\delta_y \equiv a_{c12} \delta$, $\delta_z \equiv a_{c13} \delta$, but we warn the reader that $\delta_i$ should not be confused with the component of a tensor.

We should stress that the measurement is not directly sensitive to the actual offcenterings $\mathbf\Delta$, but to the following combinations of instrumental parameters:
\begin{eqnarray}
\Delta'_x & \approx & a_{c11} \Delta_x - a_{c12} \Delta_y - a_{c13} \Delta_z \\
\Delta'_y & \approx & a_{c11} \Delta_y + 2 a_{c12} \Delta_x - a_{c23} \Delta_z \\
\Delta'_z & \approx & a_{c11} \Delta_z + 2 a_{c13} \Delta_x + a_{c23} \Delta_y.
\end{eqnarray}
Only those ``derived'' offcenterings $\Delta_{x,y,z}$ can be estimated.
Finally, note that the quadratic factors $K_{2xx}^{(j)}$ were neglected in Ref. \cite{rodriguescqg1} since they are estimated to be negligible in the real MICROSCOPE data. In this paper, we aim to be as general as possible, so that we take them into account, and set them to non-negligible values in the simulations below. Similarly, Ref. \cite{rodriguescqg1} ignored the motion of the test masses (thus, setting the time derivatives of $\mathbf \Delta$ to 0); here, we consider it since it should be taken into account in some calibration sessions.

\subsection{Modelled, estimated and ignored (combinations of) parameters} \label{ssect_calibration}

Each term of Eq. \eqref{eq_xacc} is the product of a time-varying signal by parameters related to the experiment (called ``instrumental parameters'' in the remainder of this paper). Their effects are clear from Eq. (\ref{eq_xacc}): the differential bias shifts the measured acceleration from the true acceleration; the common-mode sensitivity matrix mixes components from all axes; the differential-mode sensitivity matrix projects common-mode acceleration into the measured differential acceleration; the coupling with the angular acceleration makes the measurement depend on the satellite's motion; and ${n}^{(d)}$ adds stochastic instrumental noise to the measurement. 

The varying signals are either directly measured or are derived by analyzing tracking data, as described below and shown in Refs. \cite{hardy13b, hardycqg6}: several satellite manoeuvres (summarised in \ref{app_calibration}) allow us to discriminate between instrumental defects and estimate them separately.

Those terms come in four distinct categories and can be either corrected for, or ignored in the measurement equation (\ref{eq_xacc}):
\begin{enumerate}
\item Terms computed through models and/or on-ground restitutions:
\begin{itemize}
\item The position of the satellite as determined in the celestial reference frame J2000 \cite{robertcqg3} is expressed in the Earth's reference frame  using the appropriate rotations \cite{iers2010:lr}. From this position, the components of the gravity acceleration $\mathbf g$ and of the GGT $\left[{\mathbf T}\right]$ are computed in the Earth's reference frame \cite{metris98}; afterwards they are transformed to the J2000 celestial frame and then to the instrument frame using the attitude of the satellite \cite{robertcqg3}.
\item Like the attitude, the angular velocity ${\mathbf \Omega} $ and acceleration $\dt{\mathbf \Omega}$ result from an on-ground restitution \cite{robertcqg3}.
\item $\tilde\Gamma^{(c)}_x$, $\tilde\Gamma^{(c)}_y$ and $\tilde\Gamma^{(c)}_z$ are directly approximated by the measured quantities $\Gamma^{(c)}_x$, $\Gamma^{(c)}_y$ and $\Gamma^{(c)}_z$.
\end{itemize}
\item Terms corrected using results from calibration sessions ($\Delta'_y$, $a_{d1j}$):
  \begin{itemize}
  \item ${\Delta'_{y} }\left({{ S_{xy}}}+\dot\Omega_z \right)$
      \item $ 2\left( { a_{d11}}{ \tilde\Gamma^{(c)}_x}+{ a_{d12}}{ \tilde\Gamma^{(c)}_y}+{ a_{d13}}{ \tilde\Gamma^{(c)}_z}\right)$
  \end{itemize}
   \item Terms dependent on parameters estimated with the current session:
        \begin{itemize}
  \item $2 \tilde{b}_x^{(d)}$: in principle $\tilde{b}_x^{(d)}$ is a readily measurable constant, although it can drift in time (section \ref{sec:low-freq-vari}).
  \item $\delta_x g_x+\delta_z  g_z$:  we take advantage from the fact that $g_x$ and $ g_z$ vary at the same frequency  $f_{\rm EP}$ but in quadrature to estimate both $\delta_x$ and $\delta_z$ almost without correlation.  Since $a_{c11}\simeq 1$ and $a_{c13}\ll{}1$, $\delta_x \approx \delta$ (estimating $\delta$ being MICROSCOPE's main objective). Given the upper bound on $\delta$  from previous experiments \cite{will14}, $\delta_z=a_{c13} \delta$  is in principle not observable but a statistic over its value estimated in different sessions could give interesting indications on the quality of the experiment.
  \item $\Delta'_{x}  S_{xx} +\Delta'_{z}  S_{xz}$: the components $\Delta'_{x} $ and $\Delta'_{z} $ of the offcentering can be estimated very accurately thanks to the strong variations of $S_{xx} $ and $S_{xz} $ at $2f_{\rm EP}$
  \end{itemize}
  \item Terms neglected due to their very small magnitude \cite{hardycqg6}: 
  \begin{itemize}
  \item $\delta_y g_y$
  \item $\left( a_{c13}\Delta'_{y}+a_{c12} \Delta'_{z} \right)  S_{yz}+a_{c12} \Delta'_{y} S_{yy}+a_{c13} \Delta'_{z} S_{zz}$
\item $\left(-a_{c13}\Delta'_{y}+a_{c12}  \Delta'_{z} +2 c_{d11}\right)\dot\Omega_x+\left(2 a_{c13} \Delta'_{x}+2c_{d12} \right)\dot\Omega_y+\left(-2 a_{c12} \Delta'_{x}+2 c_{d13} \right)\dot\Omega_z$
  \end{itemize}
\end{enumerate}

\subsection{Sessions dedicated to the test of the EP}
\label{sec:case-sess-dedic}

The measurement equation \eqref{eq_xacc} is valid for all scientific sessions (calibration and EP sessions). However, it can be simplified for EP sessions where test masses are kept motionless, so that the velocities $\dt{\Delta}_x$, $\dt{\Delta}_y$, $\dt{\Delta}_z$  and the accelerations $\ddt{\Delta}_x$, $\ddt{\Delta}_y$, $\ddt{\Delta}_z$ vanish.
Further applying the corrections described in Sect. \ref{ssect_calibration}, Eq. (\ref{eq_xacc}) then simplifies to

\begin{equation}
  \label{eq:11}
  \Gamma^{(d)}_{x, {\rm corr}}=2 \tilde{b}_x^{'(d)}+\delta_x g_x+\delta_z  g_z+\Delta'_{x}  S_{xx} +\Delta'_{z}  S_{xz}+ 2 n_x^{(d)},
\end{equation}
which is the core model fitted to the data after applying the calibration parameters: in addition to the E\"otv\"os parameter  $\delta_x $ we also estimate $\delta_z $ which quantifies the amplitude of a signal proportional to $g_z$ (varying also at the $f_{\rm EP}$ frequency but in quadrature with $g_x$) and the components $\Delta'_{x}$ and $\Delta'_{z}$ of the apparent offcentering.

The next section provides more conceptual and mathematical background on the (instrumental and E\"otv\"os) parameters used in the MICROSCOPE data analysis.


\section{Parameters estimation} \label{sect_pestimation}

\subsection{Iterative weighted least square fit}

Each in-flight calibration session is dedicated to estimating one (or two) parameters and designed so that the signals sourced by those parameters have a favourable signal-to-noise ratio.
Although it is theoretically possible to cumulate all calibration sessions and estimate all parameters simultaneously from Eq. (\ref{eq_xacc}), we use the fact that they are almost independent from each other to simplify and better control the estimation process via an alternative method. We thus devised a technique to estimate each parameter iteratively, refining and updating the estimation of a given parameter using the estimation of the other parameters until some convergence criterium is reached. 

The measurement equation Eq. (\ref{eq_xacc}) is of the form $\Gamma_x^{(d)} = f(p_k, t) + n_x^{(d)}$, where $p_k$ are parameters and the time dependence is linked to measured or modeled signals $s_i(t)$. For each session, the data provides us with $\Gamma_x^{(d)}$ and all $s_i(t)$. It is then possible to perform a least-square (or similar) fit to estimate the parameters $p_k$ from a given model.

Moreover, for a given calibration session, we have a priori values $p_{k,0}$ for the parameters $p_k$, as some of them have been measured on ground, or as others may have been estimated during an earlier in-flight calibration session (Sect. \ref{ssect_calibration}). It is then possible to correct the measurement for the corresponding signals, and use an updated version of the measurement equation,
\begin{equation} \label{eq_gcorr}
\Gamma^{(d)}_{x, {\rm corr}}(t) = \Gamma_x^{(d)}(t) - f(p_{k,0}, t).
\end{equation}
This equation can finally be used to refine the estimation of some parameters $p_{ke}$, with a least-square method using the model
\begin{equation} \label{eq_corracc}
\Gamma^{(d)}_{x, {\rm corr}}(t) =  \Sigma_{ke} \frac{\partial f(p_k,t)}{\partial p_{ke}} \left(p_{ke} - p_{ke,0}\right).
\end{equation}

The actual estimation of a parameter depends on the technique used to deal with missing data (see Sect. \ref{sect_transients} for the introduction of the techniques we use to deal with them --KARMA \cite{baghi15}, M-ECM \cite{baghi16} and {\it inpainting} \cite{berge15b,pires16}). M-ECM estimates the parameters and the noise and deals with missing data all at once; however, {\it inpainting} only fills in missing data, and must be augmented by a least square estimate.

We use iteratively the {\sc Adam} (Accelerometric Data Analysis for MICROSCOPE) Fortran code to estimate parameters in the frequency domain.
Once this iterative process has converged (typically in two to three iterations), we use Eq. (\ref{eq:11}) to measure the E\"otv\"os parameter $\delta_x$ on calibrated data. The procedure is summarised in Algorithm \ref{alg_rl}.
The data analysis processes underlying {\sc Adam} are described below.

\begin{algorithm}[H]
  \caption{Iterative least-squares estimation of $N$ instrumental parameters from $M$ independent calibration sessions.}
  \label{alg_rl}
   \begin{algorithmic} 
   \State Initial prior: $\Pi = \{\pi_0(p_1), \dots, \pi_0(p_i), \dots \pi_0(p_N)\}$
   \While{not converged} 
   	 \For{$i =1$ to $M$}
   		\State Correct measurement from $i$th session with priors $\Pi_{i-1}$ (Eq. \ref{eq_gcorr})
   		\State Estimate $i$th parameter (Eq. \ref{eq_corracc}): estimator $\hat{p}_i$
	\EndFor
	\State Update prior: $\Pi = \{\hat{p}_1, \dots, \hat{p}_i, \dots \hat{p}_N\}$
   \EndWhile
   \State Estimate the E\"otv\"os parameter $\delta_x$ on calibrated data (Eq. \ref{eq:11})
   \end{algorithmic}
\end{algorithm}

\subsection{Frequency domain least square analysis: {\sc Adam}}
\label{sec:analys-freq-doma}

To allow for a least-squares analysis \cite{lupton93}, the corrected measurement equation (\ref{eq_corracc}) can be formally written
\begin{equation}
  \label{eq:1}
  {\mathbf Y} = {\left[{\mathbf A}\right]} {\boldsymbol{\theta}}+{\mathbf n},
\end{equation}
where ${\mathbf Y} $ is the vector of $N$ measurements, ${\boldsymbol{\theta} }$ is a vector of $q$ unknown parameters to estimate (e.g. E\"otv\"os parameter or test masses' offcentering),  $\left[{\mathbf A}\right]$ is the design matrix and ${\mathbf n}$ is the noise vector.
Since the models (given by Eq. \ref{eq_xacc} in the most general case and by Eq. \ref{eq:11} for EP sessions) are linear with respect to the estimated parameters, the columns of  $\left[{\mathbf A}\right]$ simply correspond to the signal associated to each parameter, sampled at the epochs of the measured acceleration.
The $N$ measurements are assumed to be regularly sampled at a frequency $f_{\rm e}$ over a duration $T$. In case of missing data the analysis takes place after reconstruction of these data using {\it inpainting} or M-ECM algorithms (Sect. \ref{sect_transients}). 

\subsubsection{Transformation of the problem from the time domain to the frequency domain}
\label{sec:from-time-domain}

In order to solve the problem in the Fourier domain, we take the Fourier transform of Eq. (\ref{eq:1}). To this aim, we make use of the Discrete Fourier Transform Operator $\left[{\mathbf F}\right]$. The DFT operator being unitary, the signal energy content is preserved by the transformation.

The new system can be simply written
\begin{equation}
  \label{eq:7}
    \hat{\mathbf Y} = {\left[\hat{\mathbf {A}}\right]} {\boldsymbol{\theta}}+\hat{\mathbf n}.
 \end{equation}
The original quantities being real, the new system can be reduced to $N$ useful real equations.
These new equations can be grouped by pair (related to real and
imaginary parts of the DFT), corresponding to frequencies
$f_k=\frac{k}{T}, k=1\cdots\lfloor\frac{N-1}{2}\rfloor$.

\begin{figure} 
\center
\includegraphics[width=0.45\textwidth]{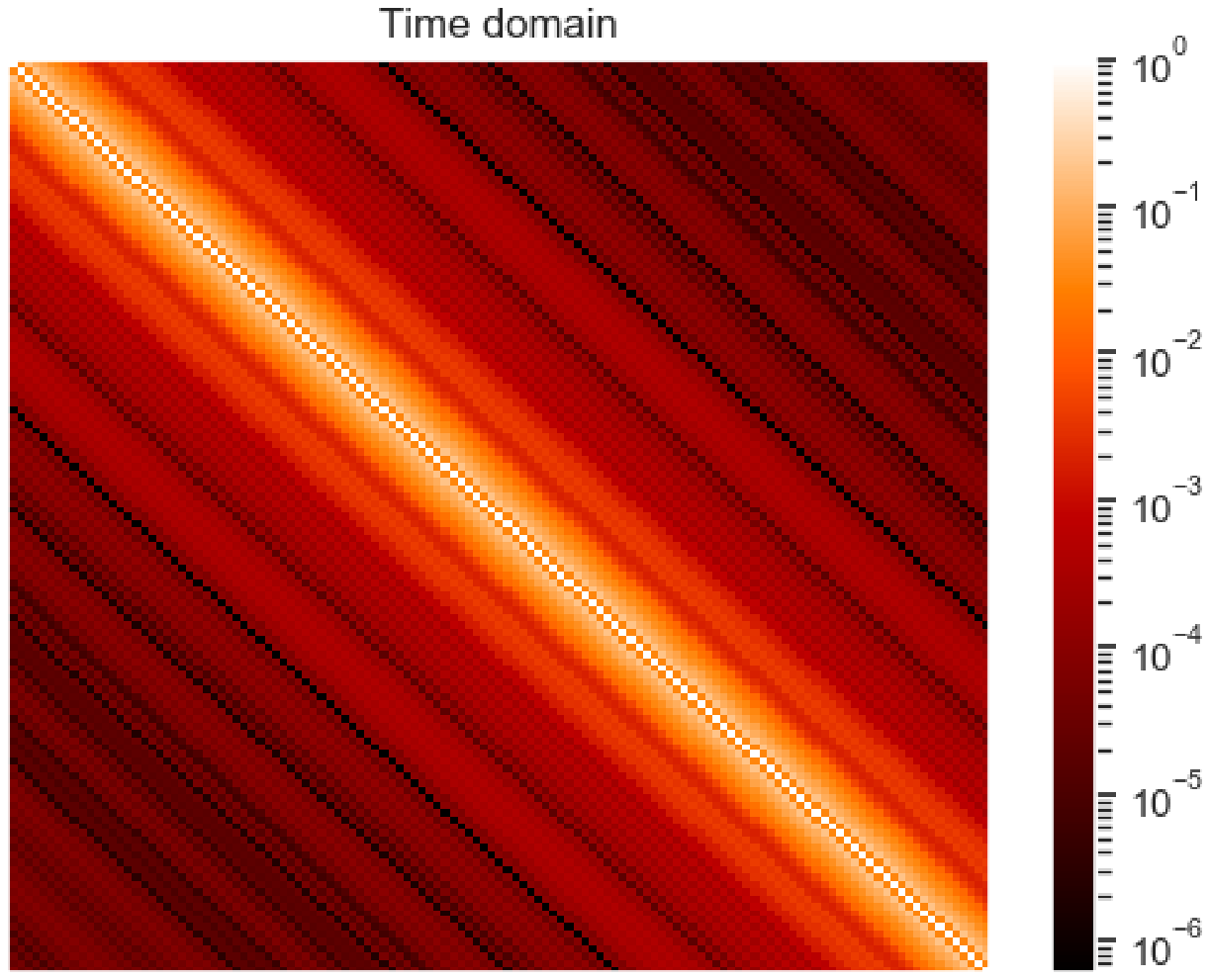}
\includegraphics[width=0.45\textwidth]{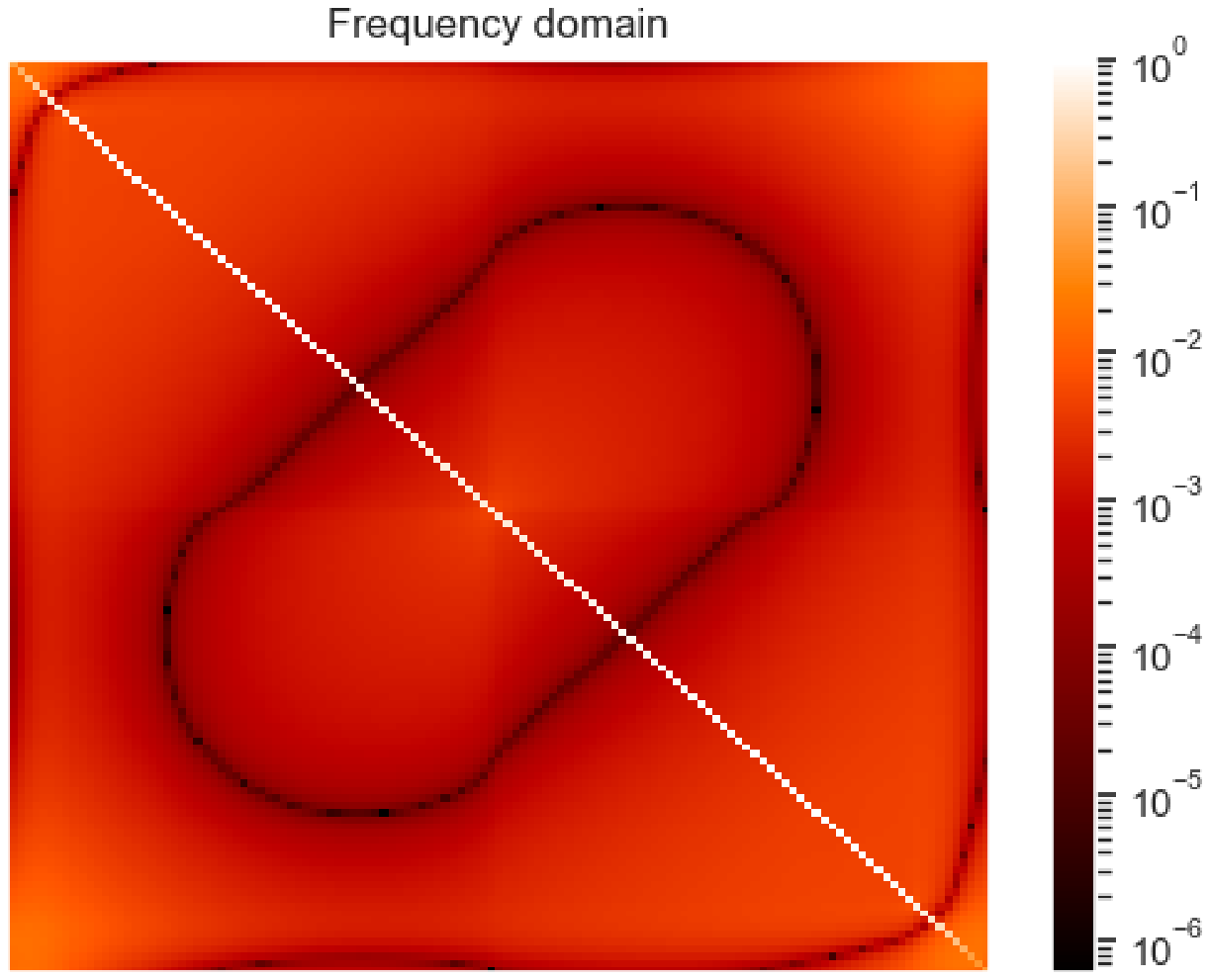}
\caption{Covariance matrix of the measured differential acceleration of an EP session.  Working in the frequency domain allows us to deal with a diagonal covariance matrix.}
\label{fig_covariance}       
\end{figure}

Moreover, the discrete Fourier transform drastically decreases the
correlations between the measurements projected in the frequency
space (Fig. \ref{fig_covariance}): the covariance matrix associated to $\left[{\mathbf F}\right]{\mathbf n}$ is diagonal dominant. This is beneficial as a diagonal weight matrix leads to a quasi-optimal solution.

\subsubsection{Weighting in the frequency domain}
\label{sec:weight-freq-doma}

Since each measurement projected in the Fourier domain can be
associated to a discrete frequency, the corresponding weight is
\begin{equation}
  \label{eq:4}
  w(f_k)=\frac{1}{\sqrt{\gamma(f_k)}},
\end{equation}
where $\gamma(f_k)$ is the Power Spectral Density (PSD) of the noise at the
frequency $f_k$. In practice, the PSD is estimated by smoothing the
residual noise in the frequency domain resulting from a first estimation and
removal of the signal. The estimation--correction--smoothing--weighting process can be iterated until
convergence; in practice, two iterations are sufficient. 

\subsubsection{Restriction to the frequency bands containing the dominant signals}
\label{sec:restr-analys-freq}

As shown in Ref. \cite{rodriguescqg1}, the MICROSCOPE mission was designed to concentrate useful signal on specific frequencies (i.e., a potential EPV signal peaks at $f_{\rm EP}$, the GGT signal at $2f_{\rm EP}$ and calibration signals at $f_{\rm cal}$). This is so true in the real data that a very simple analysis
such as synchronous detection could lead to reasonable results.
However, we use a more flexible method: we limit our least square inversion to the bands of frequency containing the relevant signals. In practice, this is equivalent to extracting a subsystem of Eq. \eqref{eq:7} by selecting the  relevant equations to get the truncated system
\begin{equation}
  \label{eq:12}
    {\left[\hat{\mathbf {A}}_r\right]} {\boldsymbol{\theta}}+\hat{\mathbf n} =\hat{\mathbf Y}_r.
\end{equation}
This trade-off between synchronous detection and the inversion of the full system brings several advantages:
\begin{itemize}
\item it is more robust than synchronous detection in case of small fluctuation of the frequencies of the signals; it is not even necessary to know precisely the value of these frequencies;
\item the choice of large enough bands containing also a substantial sample of noise allows us to compute consistent values of the goodness of fit;
\item contrary to the inversion of the full system  \eqref{eq:7} the solution of the truncated system \eqref{eq:12} is immune to possible unmodelled perturbations in frequency bands containing no useful information (especially high frequency bands). Taking these bands into account would not change the parameter estimations (because equations at different Fourier frequencies are uncorrelated) but could decrease the global goodness of the fit;
\item the number of observation equations is neatly decreased and the analyses are faster.
\end{itemize}

A final clarification is in order: the very low frequency part of the
signal, which contains in particular the zero frequency component, is not used in
the analysis. Indeed, not only does the zero frequency  contain well-known signal
such as some components of the gravity gradient but it also contains unknown
signals such as the bias of the accelerometers. All these
contributions are not separable, so that the zero frequency  is difficult to
exploit.

\subsection{Bias drift}
 \label{sec:low-freq-vari}

\begin{figure} 
\center
\includegraphics[width=0.45\textwidth]{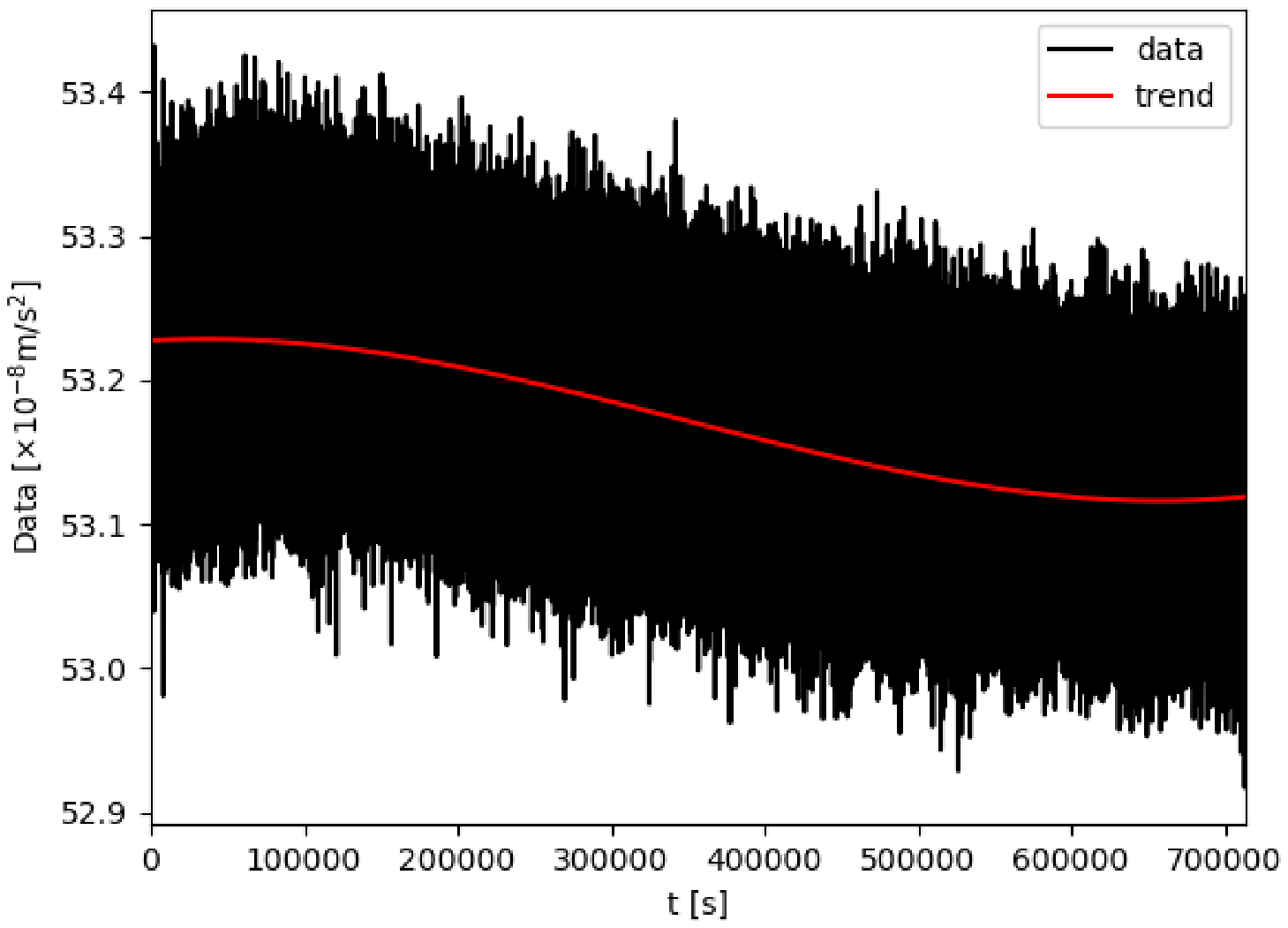}
\includegraphics[width=0.45\textwidth]{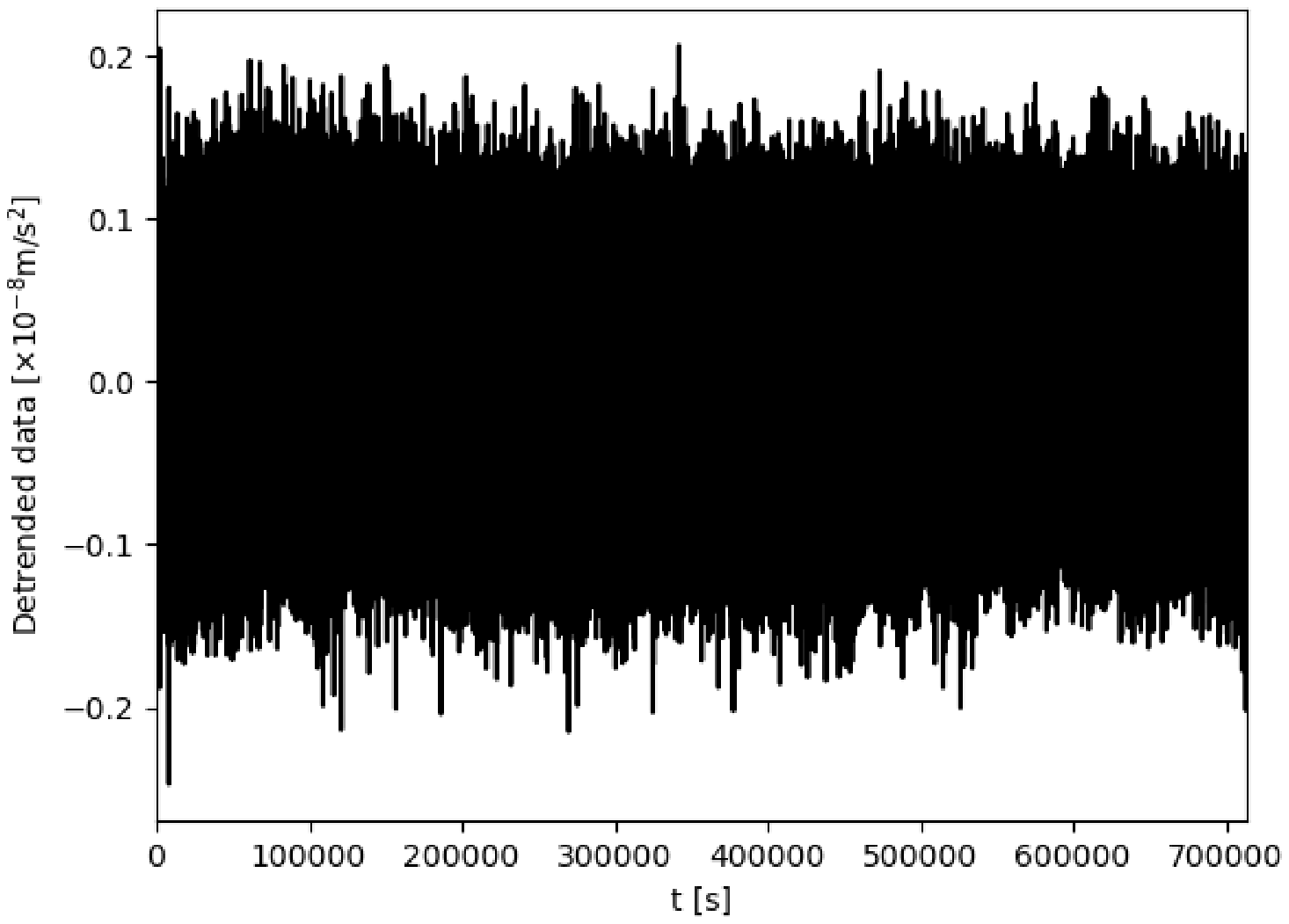}
\caption{Time evolution of the differential acceleration. Left: evidence for a polynomial-like tendency. Right: after correcting for a polynomial of degree 3.}
\label{fig_biais-varaitions-time}       
\end{figure}

\begin{figure} 
\center
\includegraphics[width=0.75\textwidth]{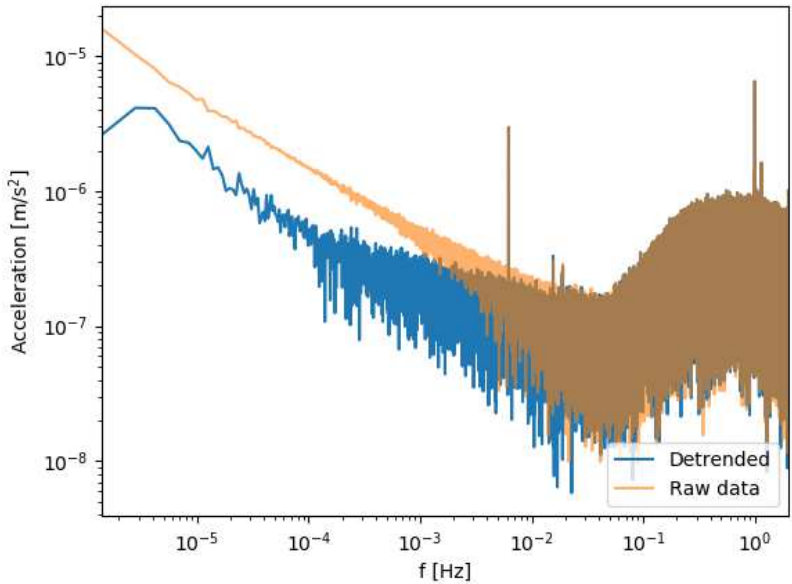}
\caption{Same as Fig. \ref{fig_biais-varaitions-time} but in  the frequency domain. As expected, the subtraction of the polynomial (blue) clearly reduces the low frequency contribution.}
\label{fig_biais-varaitions-freq}       
\end{figure}

Strictly speaking, the bias $\vv{b_0}^{(d)} $ included in
Eq. \eqref{eq_xacc} should be constant. But in practice a very low
frequency evolution can appear, in particular because of thermal variations
in the instrument (left panel of Fig. \ref{fig_biais-varaitions-time}). This evolution can be 
efficiently corrected by fitting and subtracting a polynomial (right panel of
Fig. \ref{fig_biais-varaitions-time}). As illustrated by
Fig. \ref{fig_biais-varaitions-freq}, the polynomial affects all
frequencies in a wide low frequency band; consequently, the correlation
with the EPV signal (which is concentrated at the $f_{\rm EP}$ frequency)
is very low. 
But to be sure that fitting a polynomial does not affect the signal around $f_{\rm EP}$, we take advantage of our estimation in the Fourier space to remove the contribution of these frequencies in the estimation of the polynomial. More precisely:
\begin{enumerate}
\item we first estimate simultaneously (i) the E\"otv\"os
  parameter using a narrow frequency band around $f_{\rm EP}$, (ii) the
  components $\Delta'_{x}$ and $\Delta'_{z}$ of the offcentring using
 a narrow frequency band around $2 f_{\rm EP}$, and
  (iii) the coefficients of the polynomial using a wide band from 0 to
  1 Hz but excluding the $f_{\rm EP}$ and $2 f_{\rm EP}$ bands.
\item we then use the first estimations of all these parameters
   to correct the measurements according to Eq. \eqref{eq_gcorr}; since
  this correction is performed in the time domain, we implicitly
  correct all the frequencies and in particular the polynomial is also
  applied to the $f_{\rm EP}$ and $2 f_{\rm EP}$ bands.
\item finally, we use the corrected acceleration to re-estimate the same
  parameters in the same way as in step (i).
\end{enumerate}

Another option is to correct low frequency variations with a model of temperature sensitivity. Ref. \cite{metriscqg9} shows that the substraction of a model linear with respect to the temperatures produces results similar to those obtained with a polynomial fit.

\subsection{Combining sessions}
\label{sec:analys-cumul-sess}

In order to decrease the stochastic error in the estimation of (instrumental and E\"otv\"os) parameters, it is interesting to combine measurements from several sessions with the same configuration. This combination can be done either in the time or in the frequency domain, as shown below.

\subsubsection{Time domain}
\label{sec:gath-time-doma}

Suppose that session $k$ leads to the following linear system of $N_k$ equations (of the form of Eq. \ref{eq:1}):
\begin{equation}
  \label{eq:5}
  {\left[{\mathbf A}_k\right]} {\boldsymbol{\theta}}+{\mathbf n}_k ={\mathbf Y}_k
\end{equation}
To carry on with a classical inversion in the time domain, one can
just gather the above matrices and vectors:
\begin{equation}
  \label{eq:6}
  \left[{\mathbf A}\right]=
  \begin{bmatrix}
     {\left[{\mathbf A}_1\right]} \\
 {\left[{\mathbf A}_2\right]} \\
\vdots \\
 {\left[{\mathbf A}_m\right]} 
  \end{bmatrix},
\quad
{\mathbf n}=
  \begin{bmatrix}
     {\mathbf n}_1 \\
 {\mathbf n}_2\\
\vdots \\
 {\mathbf n}_m
  \end{bmatrix},
\quad
{\mathbf Y}=
  \begin{bmatrix}
     {\mathbf Y}_1 \\
 {\mathbf Y}_2\\
\vdots \\
 {\mathbf Y}_m
  \end{bmatrix}
\end{equation}
where $m$ is the number of sessions considered. Then an appropriately weighted least-squares technique could be used to solve for the concatenated system. Given the advantages with doing an analysis in the frequency domain as outlined above, we have not used this time domain analysis.

\subsubsection{Fourier domain}
\label{sec:gath-four-doma}

First,  DFT are applied separately to the systems corresponding to
each session $k$, as described in  Sect. \ref{sec:from-time-domain},
to get equivalent systems in the frequency domain:
\begin{equation}
  \label{eq:8}
   {\left[\hat{\mathbf {A}}_k\right]} {\boldsymbol{\theta}}+\hat{\mathbf n}_k =\hat{\mathbf Y}_k.
\end{equation}
We use a weighting based on the PSD, as described in Sect. \ref{sec:weight-freq-doma}.
Finally, all considered weighted systems are gathered as shown above for the time domain case.

Like in the case of a unique session, we can use a Generalised Least-Squares regression to solve either the whole system or, more efficiently, a selected subset of equations corresponding to the desired bands of frequency. Nevertheless, the estimation process is flexible enough to allow us to take into account the drift of some parameters (e.g. offcenterings, which vary with respect to temperature).

\subsection{Error propagation} \label{ssect_errprop}

Uncertainties in estimated parameters (or in the a priori knowledge of parameters that cannot be finely estimated) propagate when correcting for them in the measured differential acceleration, which involves three effects when estimating the EPV signal: (1) residual systematic errors coming from the bias of each parameter estimate; (2) statistical errors coming from the variance of each parameter estimate and (3) statistical errors coming from the fact that we use noisy signals to calibrate the measurement, whose noise is coupled with the parameters.
The first type of errors is assessed by upper bounds given by a detailed performance analysis \cite{hardy13a, hardycqg6}. 
The second and third type of errors are discussed in this section: they can be assessed through Monte Carlo simulations or directly included in the least-square covariance matrix.

Using an iterative estimation technique involves propagating the uncertainties of the $i$th iteration estimates to the $(i+1)$th iteration. Consequently, as can be seen from Eq. (\ref{eq_gcorr}), the variance of the calibrated differential acceleration is not only that of the raw measured differential acceleration, but also involves a contribution from the (imperfect) knowledge of the corrected instrumental parameters.

Error propagation is derived rigorously in \ref{sect_appErrProp}. Eq. (\ref{eq_evalcal}) gives the expected value of the calibrated differential acceleration used to estimate the E\"otv\"os parameter when assuming that the estimates of all instrumental parameters are unbiased. Its variance is given by Eq. (\ref{eq_variance2}). Those equations can easily be generalised to the case where a given instrumental parameter is estimated given some others.

Two effects coming from uncertainties in the iterative correction of instrumental parameters (through errors in the estimated amplitudes of the signals that it contains) can be noted:
(i) statistical errors in the calibrated differential acceleration, because of the imprecise estimation of corrected parameters (during previous iterations), and
(ii) uncertainty on the calibrated differential acceleration residuals (due to the fact that we subtract noisy quantities multiplied by estimated parameters).
Therefore, we can expect that the least-square estimator of a parameter measured from the calibrated differential acceleration will be affected by statistical errors from instrumental parameters, as well as by statistical errors coming from the measurement noise, and errors on statistical errors coming from the uncertainties on the noise.

In practice, in order to rigorously propagate uncertainties in our iterative least-square estimates, Eq. (\ref{eq_variance2}) can be used as the input variance, for each parameter, at each iteration. Another possibility is to run Monte Carlo simulations that use the variance as measured from the noise of several calibrated differential accelerations created from drawings of instrumental parameters (that can be assumed normally distributed, with mean and variance defined by the previous estimations). Although more CPU- and time-consuming, this technique is more sound than using Eq. (\ref{eq_variance2}), since it does not involve ``educated guesses'' on some parameters (note that some terms appearing in Eq. (\ref{eq_variance2}) are the ``true'' values and not their estimates).

We show below that given the smallness of the uncertainty on the estimated instrumental parameters, we can actually safely ignore propagating errors on the estimation of instrumental parameters.


\section{Glitches and missing data} \label{sect_transients}

\subsection{(Masked) glitches and missing data (in practice)}

Data ``gaps'' come in two flavours: (i) glitches, where data is available but contaminated by physical (e.g. impact with a micrometeorite or a satellite crack) or measurement (e.g. an internal saturation in the instrument's servo-loop command) processes uncorrelated with testing the WEP, and (ii) telemetry losses, where data is missing during some time intervals.  
Glitches may bias the measurement or introduce spurious signals at specific frequencies, or within some frequency range \cite{bergecqg8}. Similar events are present e.g. in LIGO/Virgo data, where they hamper the detection and characterisation of candidate gravitational waves signals \cite{abbott20}, and several techniques have been developed to correct for them (see e.g. Refs \cite{cornish15, zackay19, venumadhav19, wei19, torres20}).

To counteract the direct effect of glitches on MICROSCOPE's WEP measurement, we can model and subtract them as discussed in Ref. \cite{berge19}; here, we choose to remove glitches, hence giving them the same status as telemetry losses (``missing'' data).
Missing data break the even sampling of data in time, thereby preventing us from using standard OLS in the frequency domain and, most critically creating an important spectral leakage due to noise colour, potentially burying the EPV signal in MICROSCOPE data \cite{baghi15,berge15b}.

We have developed and adapted methods to cope with such missing data, that allow us to reach the required accuracy on the EPV measurement:
\begin{itemize}
\item KARMA \cite{baghi15}: the Kalman Auto-Regressive Model Analysis is a generalized least-square technique based on an autoregressive model of the (unknown) instrumental noise, whitened with a Kalman filter.
\item M-ECM \cite{baghi16}: the Modified-Expectation-Conditional-Maximization technique allows us to maximize the likelihood of available data through the estimation of missing data by their conditional expectation, based on the circulant approximation of the complete data covariance. 
\item {\it inpainting} \cite{berge15b,pires16}: originally developed for 2D observational cosmology \cite{pires09} and 1D asteroseismology \cite{pires15}, the {\it inpainting} technique uses a sparsity prior to estimate the most probable value of missing data, therefore allowing us to use an ordinary least-square fit ({\sc Adam} --see below).
\end{itemize} 
Those techniques have already been tested on numerical simulations with the simplifying assumptions that instrumental defects were perfectly known and corrected for \cite{baghi15, baghi16, berge15b,pires16}. Note that Eq. (\ref{eq:11}) is equivalent to the form used by Refs. \cite{baghi15, baghi16}; when further correcting for offcenterings, it gives the measurement equation used by Refs. \cite{berge15b,pires16}.
Ref. \cite{touboul17} used {\sc Adam} with no correction of transients nor missing data (taking advantage of quiet and clean measurement sessions). Ref. \cite{touboul19} uses {\sc Adam} on data corrected for missing data with {\it inpainting}.

\begin{figure} 
\center
\includegraphics[width=0.65\textwidth]{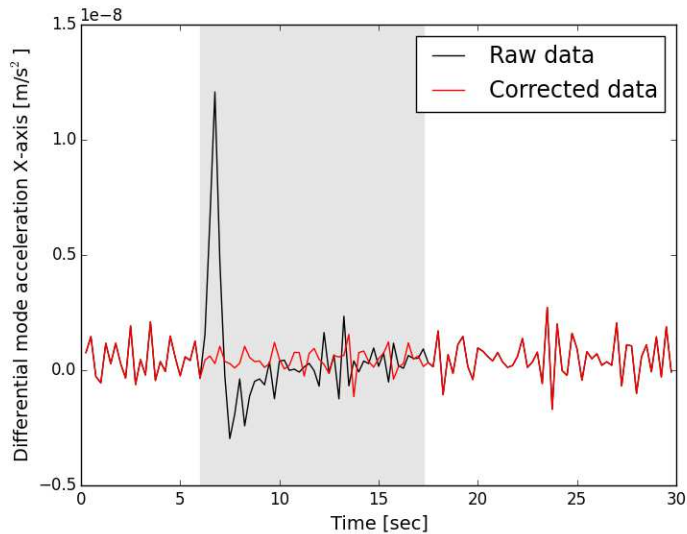}
\caption{Example of a transient and its correction. The black curve shows the raw measurement. The gray zone is the region discarded by the masking procedure. The red curve is the acceleration corrected by inpainting.}
\label{fig_gap}       
\end{figure}

To search for glitches, we use a standard recursive $\sigma$-clipping technique, in which we flag as outliers every point that deviates from the moving average of the data by more than a specific number of times (4 when defining masks on the differential-mode acceleration, or 5 when masks are defined on the common-mode acceleration) the standard deviation of the data \cite{bergecqg8}. We then mask a specific time interval after each detected outlier to make sure that the transient regime is always removed. The gray area in Fig. \ref{fig_gap} shows such a mask after an outlier to remove the corresponding glitch, visible in black, in SUSON-simulated data (see below). Additional data points are flagged by the instrument's electronics, if an internal saturation has been detected by the accelerometer digital electronics (and eventually smoothed out, thereby invisible in the acceleration data \cite{liorzoucqg2}): we decide to reject those points.
The red curve in Fig. \ref{fig_gap} shows the inpainting correction to the masked glitch (black). 

We discuss below (Sect. \ref{ssect_inpaintingLevel}) how missing data couple with MICROSCOPE instrumental parameters. This allows us to define an optimal way to deal with missing data within the iterative process described in Sect. \ref{sect_pestimation}, by considering at which level of the data processing the correction for missing data should appear.
This discussion is based on numerical simulations created by a hybrid software-hardware simulator developed by CNES (Centre National d'Etudes Spatiales) to validate MICROSCOPE's ground segment operations before flight. Those simulations correspond to worst-case scenarios, with exaggerated number of glitches and maximum allowed values for instrumental parameters. Although those simulations consider instrumental defects higher than the in-flight MICROSCOPE data, they are valuable to amplify the effect of instrumental parameters and of missing data, and allow for a pedagogical discussion hereafter (all figures in this section were produced using those simulations).

\subsection{Hybrid simulations: SUSON}

The mock data used in this section were produced by the BVSS (Banc de Validation Syst\`eme et SCAO --Drag Free and Attitude Control System Validation Bench), a CNES real time test bench which serves different purposes in project development plans. By simulating very accurately the spacecraft behavior, it aims at validating all the onboard functional chains, in particular the attitude and acceleration control. It also aims at performing all the technical and operational system tests while connected to the ground entities (operational and scientific centers).

The MICROSCOPE test bench relies on the other CNES microsatellite simulation tools, and shares many hardware and software commonalities, which gives good confidence in the product. Three benches were developed to enable the various and intensive tests performed by the project.
Each bench is constructed as follows. A Sun Microsystems station, running at 64Hz, includes various models of the spacecraft environment (Earth gravitational potential, Earth atmosphere and albedo, Sun illumination) and of its dynamics. It also includes physical models of all the embarked equipment (actuators and sensors), with major evolutions for the MICROSCOPE peculiarities (see below).
Those equipment models are coupled with a rack of interface electronic boards, emulating the real hardware communication protocol for each equipment.
A harness then connects the interface boards to a real onboard computer, on which the onboard software is running.
Finally, electrical ground support equipment simulates the ground stations and the radiofrequency link between stations and spacecraft, so that the users receive the telemetry data output by the onboard computer as in the real life.

A peculiarity of the MICROSCOPE bench is the T-SAGE\footnote{T-SAGE (Twin Space Accelerometers for Gravitation Experiments) is MICROSCOPE's instrument core \cite{liorzoucqg2}} simulation. A very representative model was developed by CNES, called SUSON. It is composed of a real Interface Control Unit (payload computer), running at 1027Hz and interfaced on one side with the main onboard computer and on the other side with a specific board simulating both the Front End Electronic Unit and the Sensing Unit \cite{liorzoucqg2}. The latter is fed with 64Hz data coming from the dynamics models of the spacecraft, so as to inject outside acceleration and gravity as sensed by the test masses. Within SUSON, the 1027Hz control loop of two tests masses is simulated with the real T-SAGE computer and software in the loop, down to the electrode level. Many parameters enable to tune the characteristic of the measurements: mass misalignments, mass centering, scale factor, quadratic scale factor, stiffness, and biases. A theoretical spectrum of T-SAGE noise is also injected. 

Another interesting aspect of this test bench is the simulation of micro-perturbations. A complementary model was developed to inject random impacts within the dynamics model, with the possibility to tune the time, direction, and momentum random distributions. The lack of flight data at the time of the first simulations campaign imposed a conservative parametrization\footnote{When compared to flight data (MICROSCOPE was not launch at the time the simulations were created, but at the time of writing, we now have in-orbit estimates of the amount of crackles), it shows that the real level of micro-perturbation is far below hypothesis.}.

Finally, we should emphasize that it was the first time a BVSS test bench was pushed so far in the simulation process. Among its notable characteristics is the wide range of timecales used: 1027Hz for the payload frequency, 64Hz for dynamics simulation, 4Hz for the onboard activation of the SCAA (Attitude and Orbit Control System) tasks, the whole process running during 10 days to obtain the final simulation outputs. 
Such a long run time is required to simulate 120-orbit sessions during which the hardware elements of the simulator ensure a very realistic simulation of the spacecraft and the possibility to validate in depth the functional chains.

\begin{figure} 
\center
\includegraphics[width=0.45\textwidth]{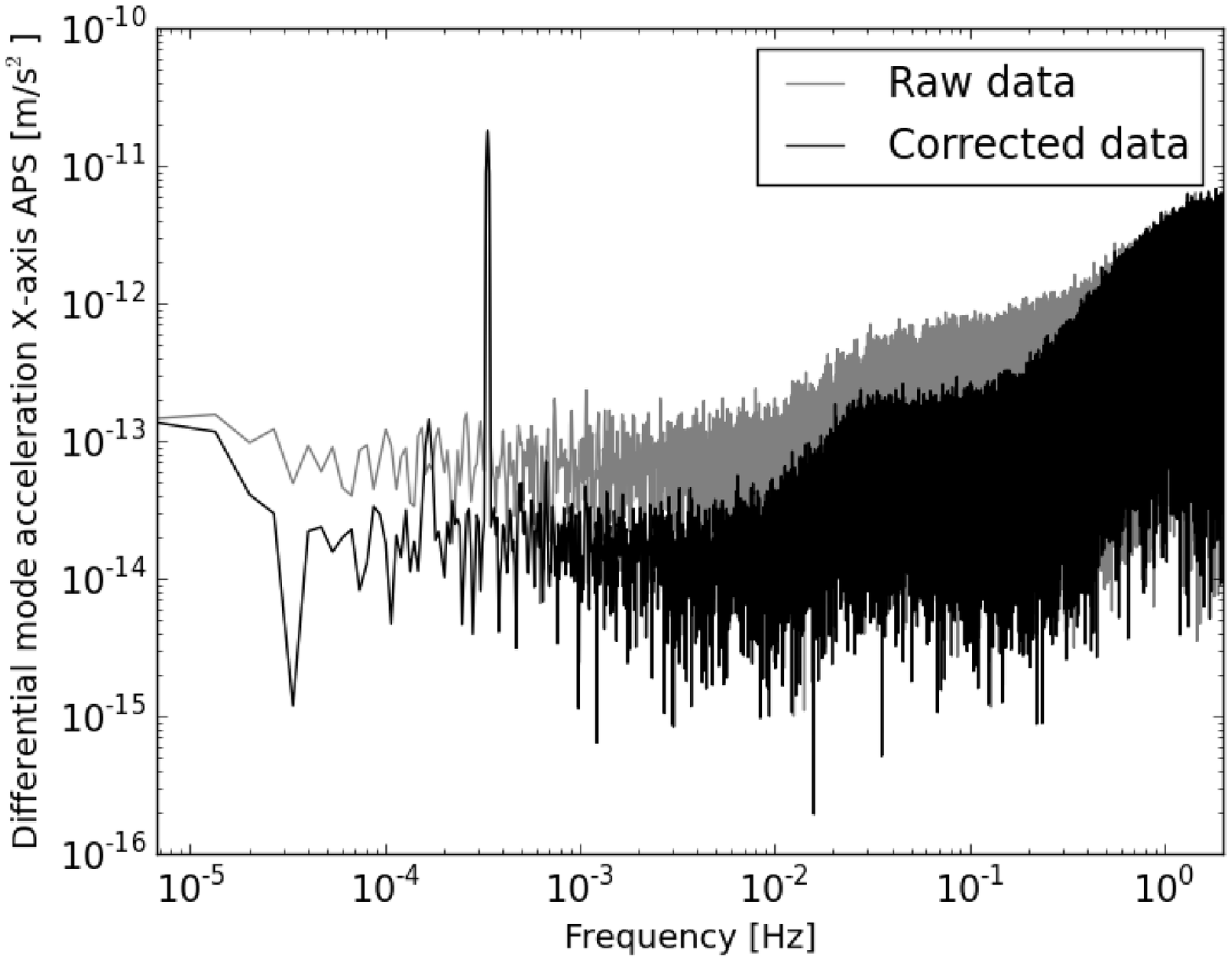}
\includegraphics[width=0.45\textwidth]{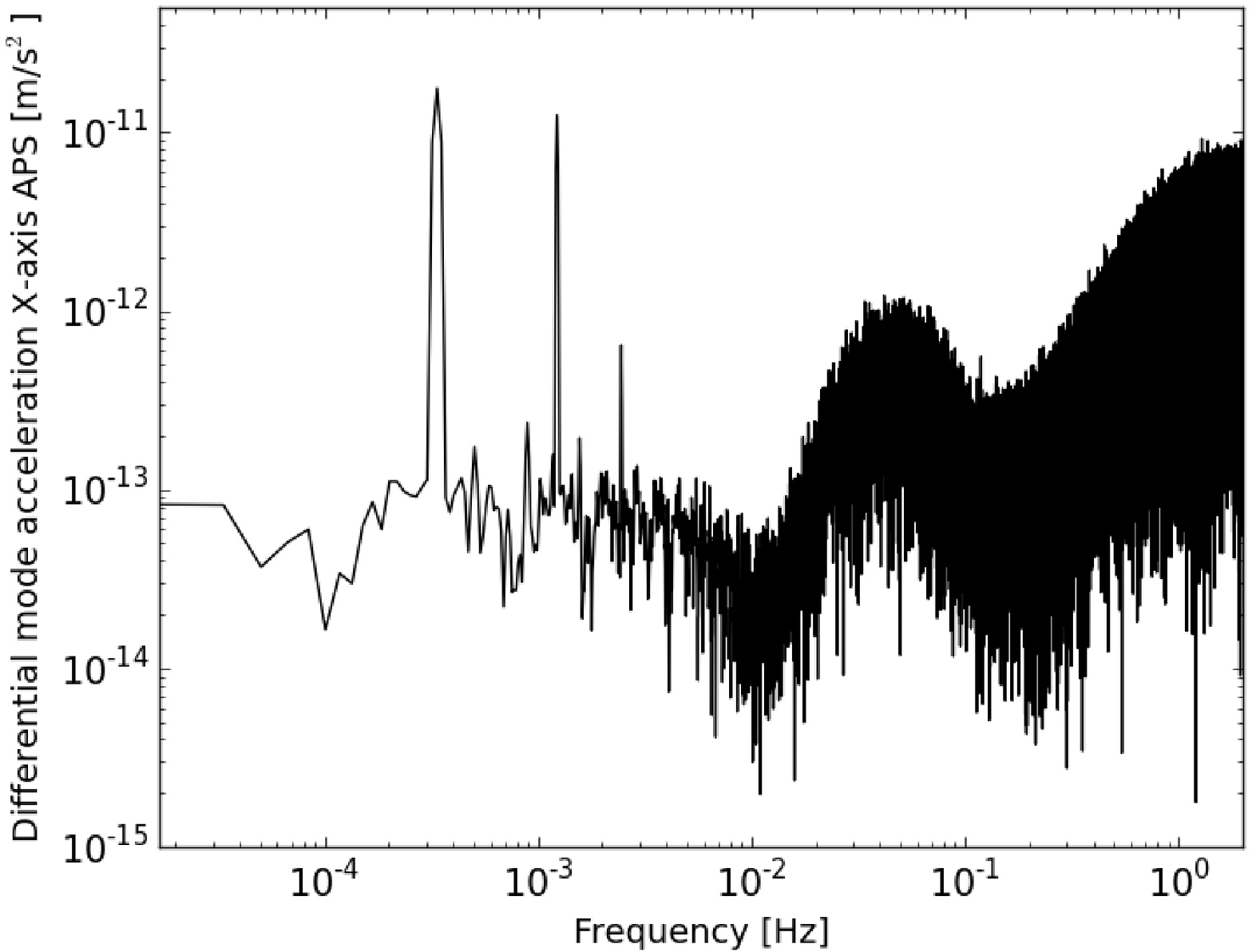}
\caption{Spectra of the simulated differential acceleration for two different measurement sessions, simulated with SUSON. Left: inertial WEP-measurement session (shown are spectra with glitches present (gray) and masked then replaced by inpainted signal (black)). Right: $\Delta_y$ calibration session.}
\label{fig_suson}       
\end{figure}

Fig. \ref{fig_suson} shows the spectra of the simulated differential acceleration for two different measurement sessions, simulated with SUSON.
The left panel shows the noise and spectral leakage induced by glitches and missing data (gray) and their correction with {\it inpainting} (black) in the frequency domain. The right panel shows the differential acceleration during a simulated $\Delta_y$ calibration session.
The bump between 0.01 Hz and 0.1 Hz is not caused by transients but is due to worst-case inaccuracies simulated in the attitude control coming from the star sensor \cite{prieur17}, and is linked to the angular velocity of the satellite. This bump is not prejudicial since it occurs at frequencies different than those of the excitations and estimated signals. Furthermore, it can be corrected for when estimating parameters in the iterative way presented in Sect. \ref{sect_pestimation}.

\subsection{Effect of the data level on glitches masking and missing data correction} \label{ssect_inpaintingLevel}

As presented in Ref. \cite{rodriguescqg4}, we use different data levels during data processing and analysis. Those relevant for our current discussion are:
\begin{itemize}
\item N1a data: raw science data sorted by inertial sensor
\item N2a data: raw science differential- and common-mode accelerations
\item N2b data: derived from N2a data after correcting for instrumental parameters. They are used to estimate the EPV signal.
\end{itemize}

We can wonder at which stage we should define the mask and correct for glitches and missing data. We show here that the answer depends on what task we aim to accomplish.

\subsubsection{Instrumental parameters estimation}

We deal with calibration at the N2a level; thus, two routes can be taken to define masks: either we mask N1a accelerations, then logically add those masks when creating N2a data, or we ignore masking N1a data and directly mask at the N2a level. One could naively think that when creating the differential acceleration, glitches would cancel out (both sensors see them in the same way, since they come from external perturbations); then, the mask needed at the N2a level should be less conservative than the logical addition of the N1a masks. However, sensors are not exactly identical, therefore leaving significant (though attenuated) imprints of glitches and other invalid data at N2a level. We find that defining the mask either at N1a level or N2a level lead to approximately the same final N2a mask, although we find that it is best to apply {\it inpainting} on N2a data level rather than on the N1a level. When applying {\it inpainting} to N1a data, reconstruction errors add up in the N2a level data, thereby providing a less optimal result. We also find that KARMA and M-ECM are more stable than {\it inpainting} with respect to the mask definition.

\subsubsection{WEP measurement: non-commutativity of inpainting and systematics subtraction}

We look for an EPV at the N2b level. Once instrumental parameters are estimated, subtracting their contribution from Eq. (\ref{eq_xacc}) can {\it a priori} be done on data still plagued by missing or invalid data, to which simple subtractions should be immune. We then have once again two routes to choose between in order to go from N2a to N2b data: either correct for instrumental parameters on already inpainted N2a data, or correct for them on raw N2a data, then apply {\it inpainting}, KARMA or M-ECM on those raw N2b data level. KARMA and M-ECM do not rely on filling missing data, but only on the mask definition. Therefore, their precision does not depend on whether we define masks at the N2b level or we create N2b level masks from N2a masks. Hence, the following discussion pertains only to methods that fill data gaps and we specialise on the {\it inpainting} case.

The instrumental parameters are coupled with the measured common-mode acceleration (through the $a_d$ parameters) and the square of the acceleration of individual sensors (through the quadratic factors $K_2$). Their subtraction will then evidently depend on whether we subtract them (i) before applying {\it inpainting} or (ii) after correcting them for invalid and missing data. In the former case, we create raw N2b data from raw N2a data, that we must eventually mask and inpaint; in the latter case, we create {\it inpainted} N2b data from {\it inpainted} N2a data. The guiding principle to choose between those two possibilities is to maximise the consistency between N2a data (inpainted or raw) and systematic/instrumental effects. 

\begin{figure} 
\center
\includegraphics[width=0.45\textwidth]{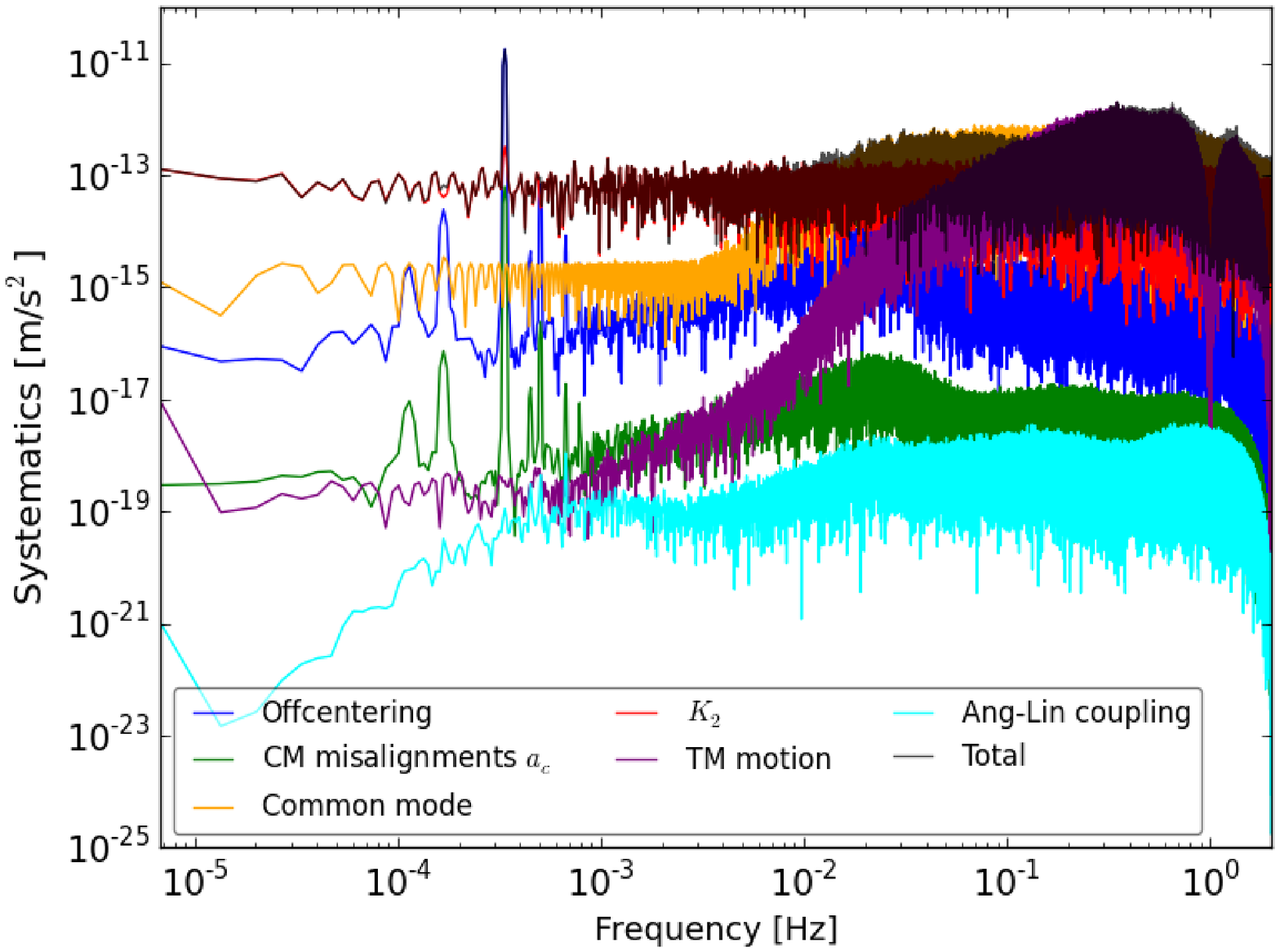}
\includegraphics[width=0.45\textwidth]{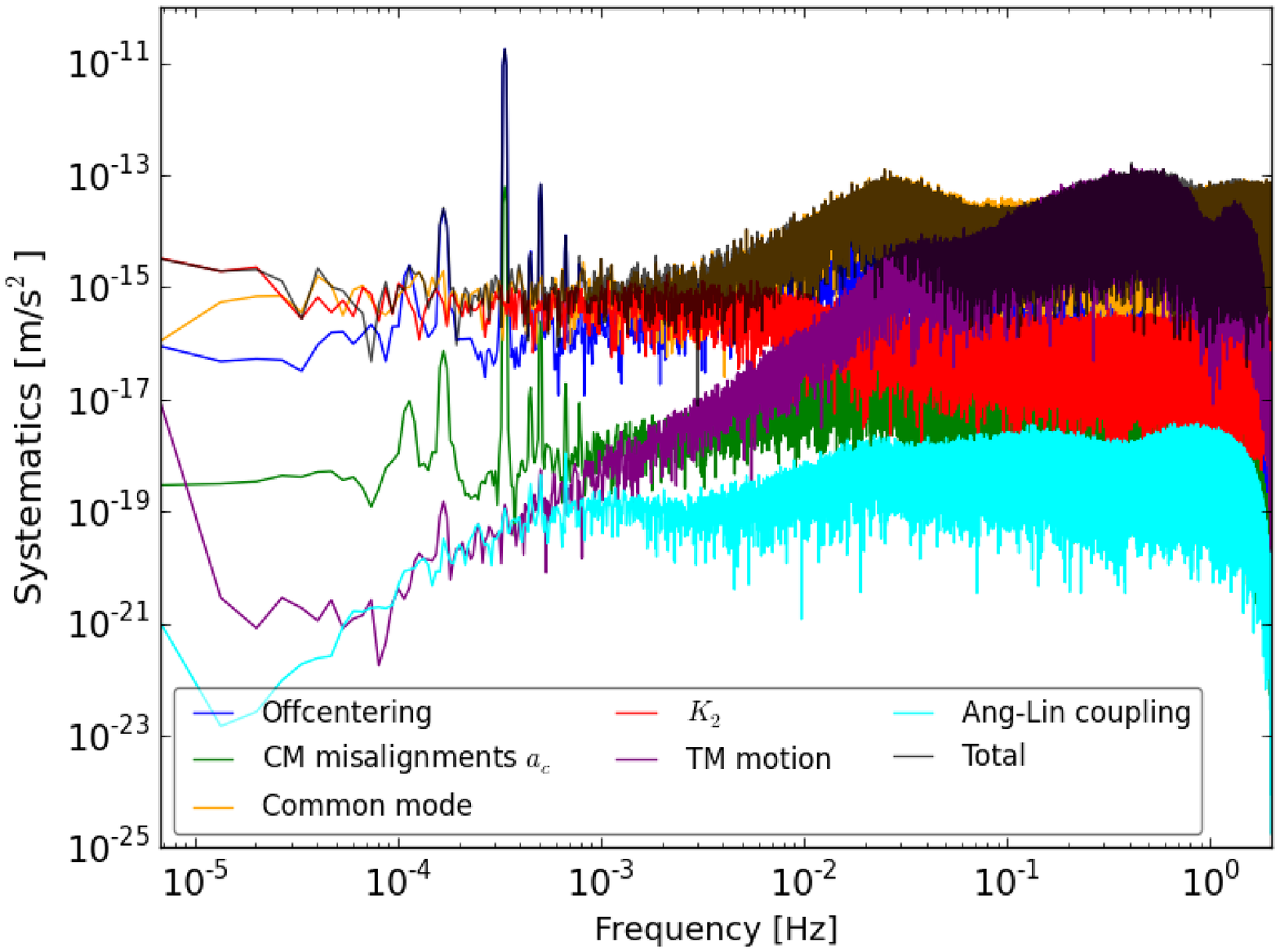}
\caption{Systematics and instrumental parameters contributions on a SUSON simulation. Left panel: with no correction of missing data. Right panel: after correction of missing data. See main text for the description of each curve.}
\label{fig_systs}       
\end{figure}

We start by computing the systematic/instrumental parameters that are to be subtracted from N2a data.
Fig. \ref{fig_systs} shows the different contributors to systematic effects when the common-mode and individual accelerations are not inpainted (left panel) and are inpainted (right panel). In those figures, the blue curve shows the main effect of the Earth gravity gradient combined with the offcentering of the test masses; the green curve shows the effect of the common-mode misalignments $a_{cij}$; the purple curve shows the effect of the test masses' motion and satellite's attitude; the orange curve shows the effect of the projection of the common-mode accelerations; the cyan curve shows the effect of the angular to linear couplings; the red curve shows the effect of the quadratic factors; the transparent black curve is the total systematics.

The main deterministic systematics come from the coupling between the test-masses offcentering and the gravity gradient. The 2$f_{\rm EP}$ line can be clearly seen, but the $f_{\rm EP}$ peak is strong enough to mimic an EPV violation of about $10^{-14}$ m/s$^2$ if not corrected for.
Furthermore, the spectral leakage from the mean of the quadratic factor dominates at almost all frequencies. It is responsible for the flat spectrum at low frequency, at a level of $10^{-13}$ m/s$^2$ before inpainting and $10^{-15}$ m/s$^2$ after inpainting. This plateau can be seen clearly on the N2a differential acceleration with the gray curve of the left panel of Fig. \ref{fig_suson}.

\begin{figure} 
\center
\includegraphics[width=0.45\textwidth]{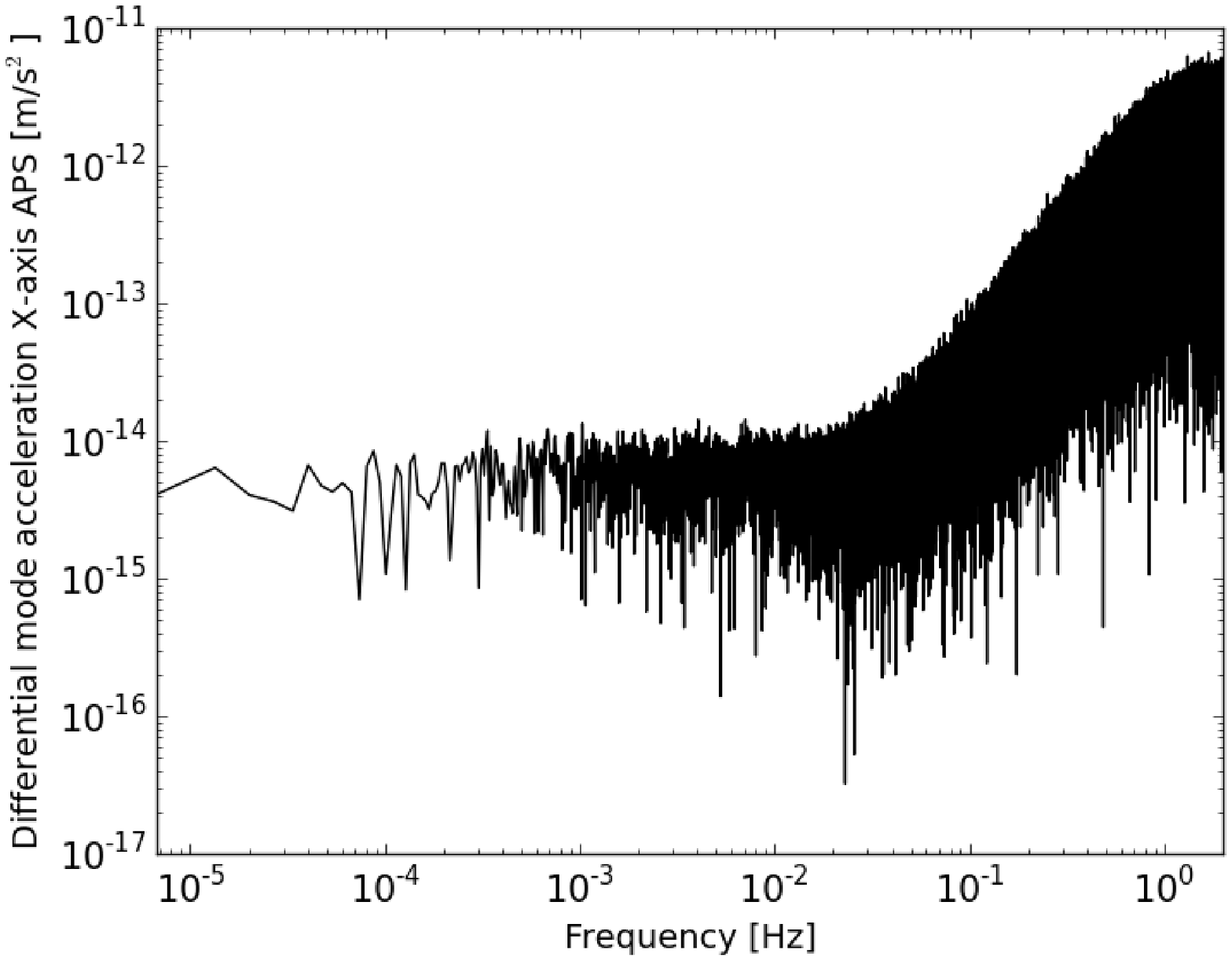}
\includegraphics[width=0.45\textwidth]{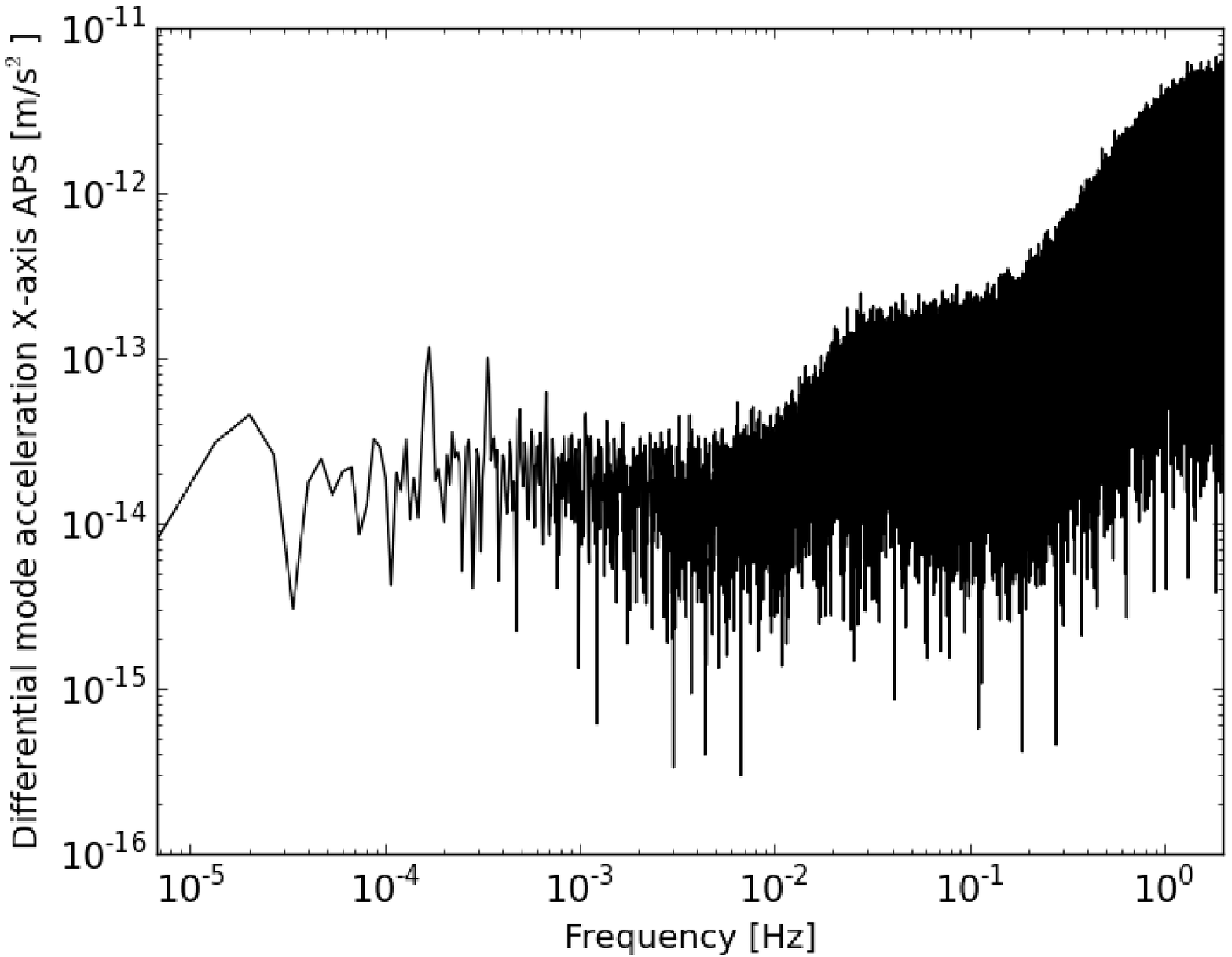}
\caption{Left panel: inpainted N2b differential acceleration created from raw N2a differential acceleration and systematics. Right panel: N2b differential acceleration created from inpainted N2a differential acceleration and systematics.}
\label{fig_n2b_appab}       
\end{figure}

The two options to create N2b data corrected for invalid and missing data can be seen graphically as (i) subtracting the black curve (at the data level) of Fig. \ref{fig_systs}'s left panel to Fig. \ref{fig_suson}'s gray curve, then mask the resulting data and apply {\it inpainting} to it or (ii) just subtracting the black curve (at the data level) of Fig. \ref{fig_systs}'s right panel to Fig. \ref{fig_suson}'s black curve. The resulting N2b spectra are shown in Fig. \ref{fig_n2b_appab}.

The differences are striking. On the one hand, subtracting raw systematics to raw N2a differential accelerations, then mask and inpaint them (left panel of Fig. \ref{fig_n2b_appab}), provides a very clean N2b differential acceleration, with no significant deterministic signals. On the other hand, subtracting inpainted systematics to inpainted N2a differential acceleration provides a poor N2b acceleration (right panel of Fig. \ref{fig_n2b_appab}), with a higher low-frequency noise and still plagued by deterministic signals, with peaks at $f_{\rm EP}$ and 2$f_{\rm EP}$ and the ``attitude'' bump about $10^{-2}$Hz (right panel of Fig. \ref{fig_suson}): although this bump is not problematic to estimate an EPV signal, the peak at $f_{\rm EP}$ would lead us to incorrectly conclude for the existence of an EPV.

\begin{figure} 
\center
\includegraphics[width=0.65\textwidth]{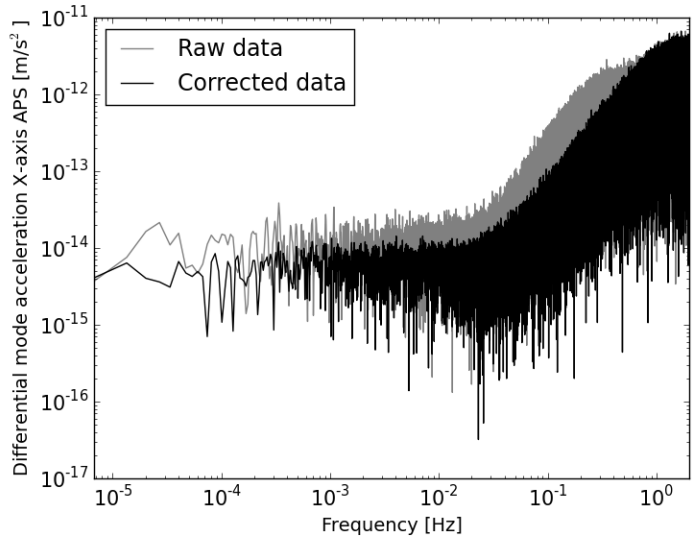}
\caption{Spectrum of N2b level differential acceleration for the inertial session used to estimate the EPV in SUSON simulations, before (grey) and after (black) glitches and missing data correction. This spectrum can be compared with that shown (in black) in Fig. \ref{fig_suson} before correcting for instrumental parameters.}
\label{fig_n2b}       
\end{figure}

This can be explained in the following way. Systematics and instrumental parameters are coupled with the common-mode acceleration and with the individual test mass squared accelerations; those are to be subtracted from the (N2a) differential-mode acceleration. Although there seems to be four different accelerations involved in this process, there are actually only two (whose sum and difference make up the common-mode and differential-mode accelerations), and hence all accelerations are significantly correlated (up to the differences of both sensor's transfer function). In particular, transients are seen for both test masses (albeit not in exactly the same way due to their different transfer functions). 
Therefore, when subtracting raw systematics to raw N2a differential acceleration, we consistently remove most of the invalid data, and we almost get clean N2b data (see the gray spectrum in Fig. \ref{fig_n2b}); masking the remaining outliers and inpainting them then improves the spectrum (to obtain the clean N2b data of the black curve of Fig. \ref{fig_n2b}).

In contrast, when masking and inpainting N2a data and N1a data (needed to subtract the quadratic factors, that couple to the individual accelerations --see Eq. \ref{eq_xacc}), {\it inpainting} introduces small errors (although not visible on individual spectra) that break the consistency between the four accelerations involved. In particular, if some bias (that can be different for the differential-mode and common-mode acceleration) is introduced by {\it inpainting} in the inpainted deterministic signals, their subtraction will be imperfect, resulting in an incorrect N2b differential acceleration. Such biases are the likely cause of the remaining peaks in the spectrum of the lower panel of Fig. \ref{fig_n2b_appab}.
Furthermore, if transients are periodic, those inconsistent errors will create a periodic pattern at the N2b differential acceleration, that can plague the {\it inpainting} reconstruction, thereby introducing an artefact at the frequency of this pattern. Such an artefact can increase the appearance of a sub-optimally subtracted deterministic systematics.

To summarise this discussion: the invalid/missing data correction (using {\it inpainting}) and systematics substraction do not commute. Therefore, we first create raw N2b-level differential accelerations from raw N2a accelerations and raw systematics, before masking them and filling their gaps. This allows us to apply {\it inpainting} only once (thus using it in the regime explored in \cite{berge15b,pires16} and minimising the bias observed in \cite{pires16}), and to take advantage of the correlations between all the accelerations involved in the process.

\section{Validating iterative least squares through numerical simulations} \label{sect_pysimula}

Although the SUSON simulations used above are sufficient to investigate qualitatively the impact of instrumental parameters and the effect of missing data, they are not precise enough to test our full pipeline down to the precision required for MICROSCOPE. This is because they were not designed for such a high-precision task, but rather to make sure that all ground-segment systems worked as expected. Consequently, we developed Monte Carlo numerical simulations dedicated to test the data analysis technique presented in Sect. \ref{sect_pestimation}, to show the internal consistency of our iterative least-square estimations in the absence of data gaps. We already showed that KARMA, M-ECM and {\it inpainting} allow us to correctly deal with missing data when estimating (in a non-iterative way) parameters \cite{baghi15,baghi16,berge15b,pires16}, thereby allowing us to ignore missing data in the remainder of the paper.

We assess the accuracy and precision of (i) ADAM's least-squares estimator and of (ii) the iterative calibration of instrumental parameters.
We investigate the first point on focusing on how well {\sc Adam} recovers the $\Delta'_x$ offcentering input in the simulations, when all other parameters are either ignored or perfectly corrected for. The second point is investigated by simulating 200 sets of calibration sessions and running our iterative least-square on them. In particular, such Monte Carlo simulations allow us to easily look into error propagation.
All the simulations use the same instrumental parameters and noise characteristics, each of them with its own noise realisation (simulated from the power-law noise power spectral density discussed in Ref. \cite{touboul09}), so that the dispersion of {\sc Adam}'s outputs are solely statistical and representative of the data analysis process.

\subsection{Simulations}

We use {\sc Simula}, a numerical fortran simulation code developed specifically for MICROSCOPE, to simulate the acceleration measured by both test masses. 
The simulator takes into account instrumental parameters (Eq. \ref{eq_xacc}) the satellite's orbit and attitude (either measured from real data or simulated by any external orbit simulator --e.g. GINS \cite{bourda08} for the orbit) and computes the Earth GGT in the instrument's frame. 
Finally, the motion of each test mass is simulated individually along its orbit (see Ref. \cite{hardy13b, rodriguescqg1} for the acceleration of individual test masses). 
Measurement errors and further systematic effects (such as, but not limited to, attitude errors, datation errors, drifts, periodic perturbations, missing data, scale factor stability, stability of coupling between axes) can be taken into account, but the electronic servo-loop used to control the test masses cannot easily be simulated, and is therefore ignored in this paper. 
Although we simulate the effect of instrumental parameters (misalignments, scale factors, couplings\dots), we ignore systematic effects not linked to the instrument itself (e.g. drag-free residual or attitude control imperfections) as well as transients and missing data. However, we take care to define all calibration sessions as they are performed in reality (i.e. with the same duration and manoeuvres characteristics, see Ref. \cite{hardycqg6}). Using those simplified simulations allows us to check that the least-squares estimator is unbiased.

The values of the parameters that can be estimated are the same as in the SUSON simulations of Sect. \ref{sect_transients}, and are listed in the second column of Table \ref{tab_errprop}. We use exaggerated values, so that the instrumental defects' effect is most easily visible. Besides instrumental parameters, we simulate an EPV $\delta=10^{-14}$. Our goal is thus to recover all instrumental parameters, and estimate the EPV with a $7\times{}10^{-15}$ precision (given the length and noise level of the corresponding simulated session) after correcting them from the simulated differential acceleration.

\subsection{Results} \label{sect_analysis}

\subsubsection{{\sc Adam} estimates' bias and variance}

\begin{figure} 
\center
\includegraphics[width=0.45\textwidth]{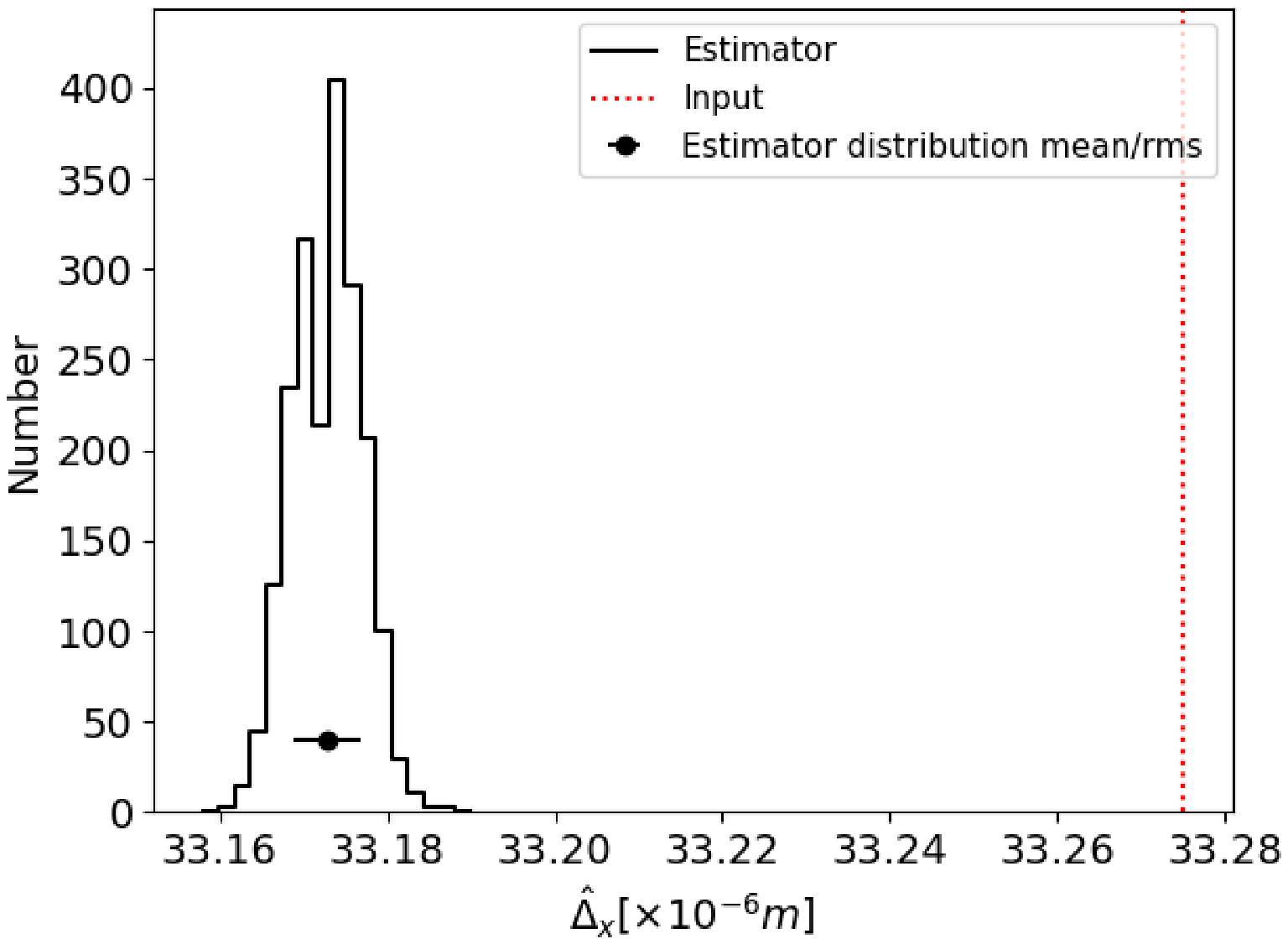}
\includegraphics[width=0.45\textwidth]{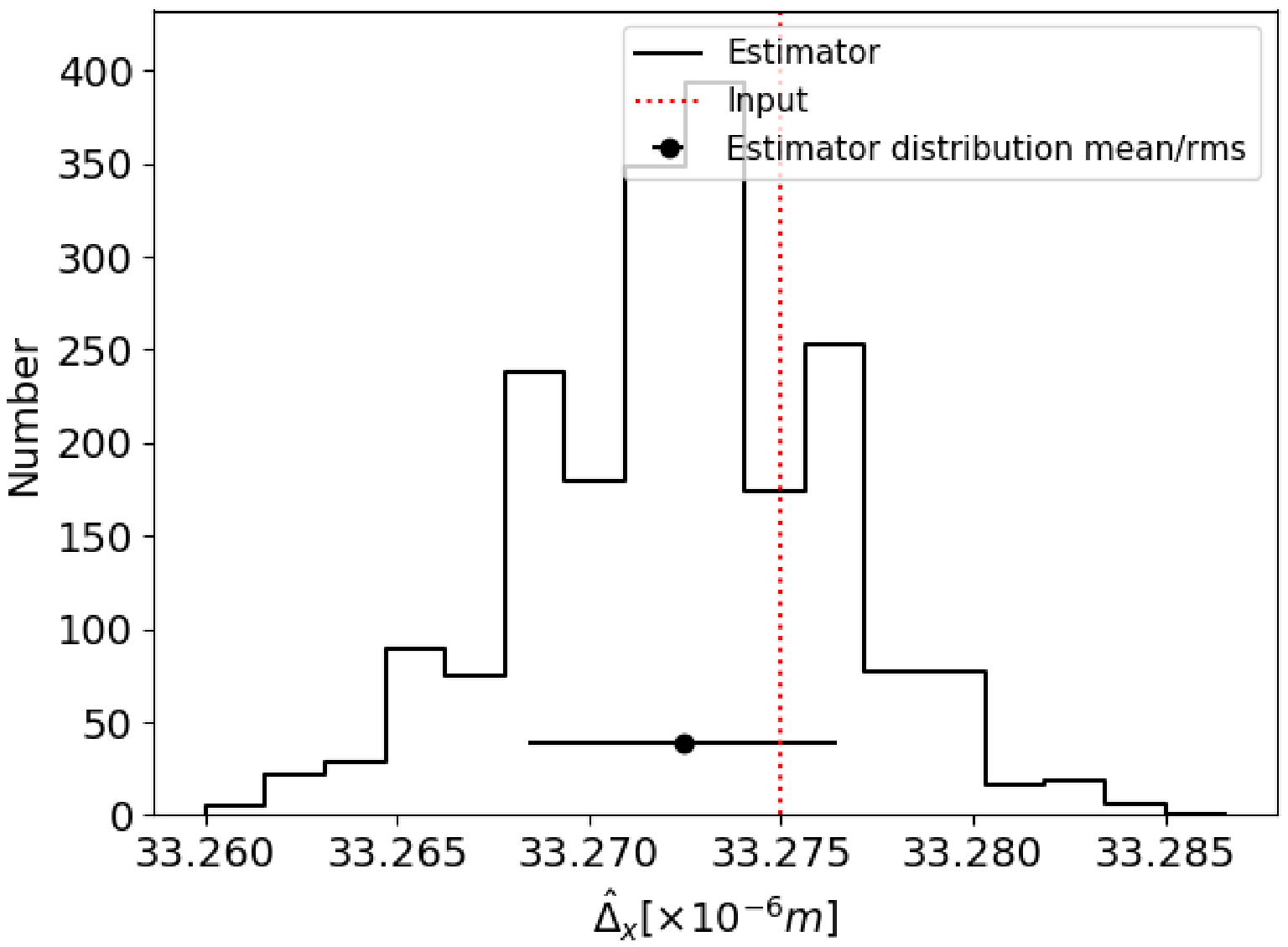}
\caption{Distribution of $\hat\Delta'_x$ estimated with {\sc adam}, from 2000 simulations. The red dotted vertical line is the value to be recovered. The black dot and error bar show the mean and rms of the distribution. Left: without correcting for other parameters. Right: with perfect prior knowledge of other parameters}
\label{fig_m2p1}       
\end{figure}

We first check that the {\sc Adam} least-square estimates are not intrinsically biased. To this aim, we measure the $\Delta'_x$ offcentering, first with no prior on other instrumental parameters, i.e. without correcting for them (which is equivalent to the first iteration of the iterative calibration).
The left panel of Fig. \ref{fig_m2p1} shows the distribution of the $\hat\Delta'_x$ estimator, based on 2000 simulations; the black dot and associated error bar show the mean and rms of the distribution; the input offcentering (as should be recovered by {\sc Adam}) is shown by the red dotted vertical line on the right hand side of the figure.
This estimation is clearly biased; although the bias is small ($\approx$ 1\%), it is highly significant.

The observed bias is due to other uncorrected instrumental parameters affecting the measurement, which justifies the use of an iterative technique.

To make sure of this assertion, we now consider the estimation of $\Delta'_x$ when other instrumental parameters are perfectly known, and can therefore be corrected for in the measured differential acceleration before estimating $\Delta'_x$. The right panel of Fig. \ref{fig_m2p1} shows the distribution of the $\hat\Delta'_x$ estimators in this case: {\sc adam}'s estimate is now unbiased, proving that the bias observed above was only due to ignoring the effect of other parameters. We can therefore claim that it is essential to iterate across instrumental parameters when measuring them and that the iterative {\sc Adam}'s estimation is unbiased. Nevertheless, we will show below that the E\"otv\"os parameter estimation is insensitive to small biases on instrumental parameters, and therefore does not require an iterative process.

\begin{figure} 
\center
\includegraphics[width=0.65\textwidth]{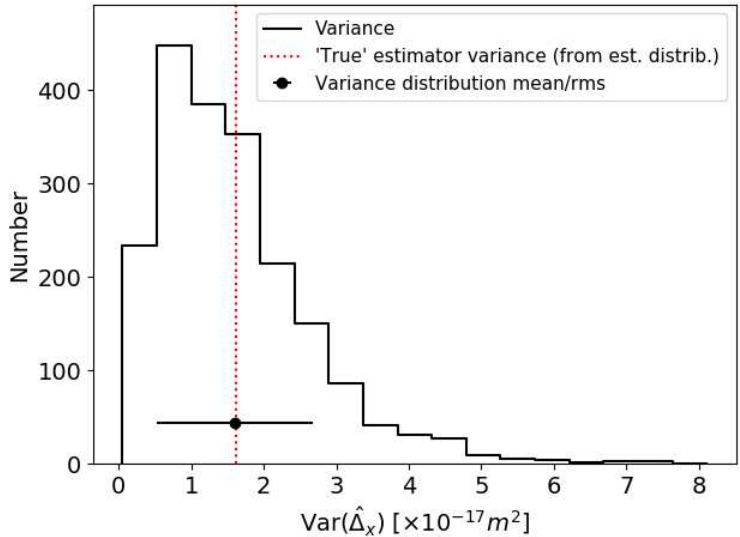}
\caption{Distribution of Var($\hat\Delta'_x$) estimated with {\sc adam}. The red dotted vertical line is the ``true'' variance as computed from the distribution of the estimator (Fig. \ref{fig_m2p1}). The black dot and error bar show the mean and rms of the estimated variance distribution.}
\label{fig_m2p2_var}       
\end{figure}

We show the distribution of the estimators' variance in Fig. \ref{fig_m2p2_var}. We can first notice that {\sc adam}'s estimated variance is distributed as a $\chi^2$-distribution, as expected for a least-square estimator's variance. In this figure, the red dotted vertical line is the ``true'' variance, as computed from the estimator's distribution (Fig. \ref{fig_m2p1}).
The black dot and associated error bars give the mean and rms of the variance of {\sc Adam}'s individual estimator. Since the ``true'' variance is within those error bars, we can conclude that the variance provided by {\sc Adam} for each estimation is reliable. This is particularly important in the MICROSCOPE landscape, and we can safely conclude that the upper bounds statistical errors on the E\"otv\"os parameter provided in Refs. \cite{touboul17,touboul19} are correct.

We can then safely conclude that {\sc adam} is intrinsically unbiased and provides correct error bars.
We checked that our conclusions about $\Delta'_x$ still hold for different values of $\Delta'_x$, as well as for other instrumental parameters.

\subsubsection{Full iterative calibration}

\begin{table}[t]
\caption{\label{tab_errprop}%
Results of the Monte Carlo simulations: measured value of instrumental parameter at each iteration. For each parameter, the first line provides the results when propagating errors from one iteration to the next; the results listed in the second line ignore the error propagation. 
The second column gives the input values with the precision required to meet MICROSCOPE's objective in brackets.}
\begin{center}
\resizebox{\textwidth}{!}{%
\begin{tabular}{c|c|ccccc}
& Expected & Iter 1 & Iter 2 & Iter 3 & Iter 4 & Iter 5  \\
\hline
$\Delta'_x$ [$\mu$m] & 33.275 & $33.172\pm{}0.025$ & $33.303\pm{}0.023$ & $33.266\pm{}0.023$ & $33.264\pm{}0.024$ & $33.269\pm{}0.028$ \\
 & (0.1) & $33.171\pm{}0.025$ & $33.301\pm{}0.026$ & $33.269\pm{}0.026$ & $33.267\pm{}0.028$ & $33.271\pm{}0.029$  \\
 \hline
$\Delta'_y$ [$\mu$m] & -16.900 & $-17.552\pm{}0.010$ & $-17.268\pm{}0.012$ & $-17.368\pm{}0.011$ & $-17.368\pm{}0.013$ & $-17.367\pm{}0.013$ \\
 &  (2) & $-17.550\pm{}0.011$ & $-17.265\pm{}0.011$ & $-17.368\pm{}0.012$ & $-17.367\pm{}0.013$ & $-17.367\pm{}0.012$  \\
 \hline
$\Delta'_z$ [$\mu$m] & -26.559 & $-26.561\pm{}0.026$ & $-26.559\pm{}0.027$ & $-26.559\pm{}0.027$ & $-26.557\pm{}0.023$ & $-26.554\pm{}0.023$  \\
 & (0.1) & $-26.557\pm{}0.021$ & $-26.557\pm{}0.026$ & $-26.556\pm{}0.025$ & $-26.556\pm{}0.024$ & $-26.557\pm{}0.027$ \\
 \hline
$a'_{d11}$ & 9.877 & $9.8866\pm{}0.0004$ & $9.8865\pm{}0.0004$ & $9.8865\pm{}0.0003$ & $9.8865\pm{}0.0004$ & $9.8866\pm{}0.0004$ \\
$[\times{}10^{-3}]$ & (0.15) & $9.8866\pm{}0.0004$ & $9.8865\pm{}0.0004$ & $9.8865\pm{}0.0003$ & $9.8865\pm{}0.0003$ & $9.8865\pm{}0.0004$  \\
 \hline
$a_{d12}$ & 1.5983 & $1.5941\pm{}0.0004$ & $1.5940\pm{}0.0004$ & $1.5941\pm{}0.0004$ & $1.5940\pm{}0.0003$ & $1.5941\pm{}0.0003$ \\
$[\times{}10^{-3}]$ & (0.05) & $1.5940\pm{}0.0004$ & $1.5940\pm{}0.0004$ & $1.5941\pm{}0.0004$ & $1.5940\pm{}0.0004$ & $1.5940\pm{}0.0003$  \\
 \hline
$a_{d13}$ & -1.4026 & $-1.4067\pm{}0.0004$ & $-1.4068\pm{}0.0003$ & $-1.4068\pm{}0.0003$ & $-1.4067\pm{}0.0003$ & $-1.4068\pm{}0.0004$ \\
$[\times{}10^{-3}]$ & (0.05) & $-1.4068\pm{}0.0003$ & $-1.4068\pm{}0.0004$ & $-1.4068\pm{}0.0004$ & $-1.4068\pm{}0.0004$ & $-1.4068\pm{}0.0004$  \\
 \hline
$a_{c12}$ & 2.99 & $4.39\pm{}0.05$ & $2.84\pm{}0.05$ & $2.83\pm{}0.05$ & $2.83\pm{}0.04$ & $2.83\pm{}0.05$ \\
$[\times{}10^{-3}]$ & (0.9) & $4.39\pm{}0.04$ & $2.83\pm{}0.04$ & $2.82\pm{}0.05$ & $2.83\pm{}0.04$ & $2.84\pm{}0.04$  \\
 \hline
$a_{c13}$ & -3.75 & $-4.95\pm{}0.04$ & $-3.55\pm{}0.05$ & $-3.55\pm{}0.05$ & $-3.56\pm{}0.05$ & $-3.56\pm{}0.04$ \\
$[\times{}10^{-3}]$ & (0.9) & $-4.94\pm{}0.05$ & $-3.55\pm{}0.05$ & $-3.54\pm{}0.05$ & $-3.56\pm{}0.04$ & $-3.56\pm{}0.05$  \\
 \hline
$K_{2d}$ [s$^2$/m] & 5612.6 & $5746.6 \pm{}1.2$ & $5746.7 \pm{}1.2$ & $5746.7 \pm{}1.2$ & $5746.6 \pm{}1.1$ & $5746.7 \pm{}1.2$ \\
 & (250) & $5746.7 \pm{}1.2$  & $5746.7 \pm{}1.1$ & $5746.7 \pm{}1.2$ & $5746.9 \pm{}1.1$ & $5746.8 \pm{}1.2$ \\
 \hline
$\delta$ [$\times{}10^{-15}$] & 10 & $7.95\pm{}7.39$ & $10.31\pm{}8.07$ & $11.63\pm{}7.97$ & $11.29\pm{}6.58$ & $9.52\pm{}6.68$ \\
 & (7) & $8.45\pm{}7.48$ & $9.58\pm{}7.00$ & $10.81\pm{}7.80$ & $9.50\pm{}7.34$ & $8.94\pm{}7.30$ \\
\hline
\end{tabular}}
\end{center}
\end{table}

After focusing on the $\Delta'_x$ estimation above, we now consider all parameters to mimic a real in-flight iterative calibration. We estimate the parameters listed in the first column of Table \ref{tab_errprop}, with input values listed in the second column; the required combined accuracy and precision of the estimation of each parameter is given in Table \ref{table_calib}.
The remaining columns list the values measured for each parameter at each iteration of the process. For each parameter, the first line gives the value estimated when considering error propagation from one iteration to the other; in the second line, we ignore error propagation and use for the $i$th iteration the best fit of each parameter obtained at the $(i-1)$th iteration. 
The best fit and uncertainty for each parameter are computed as the mean and rms of the distribution of 100 estimates from 100 simulations for each parameter and iteration. For the estimation of a given parameter, all simulations share the same deterministic signals but have a different realisation of the instrumental noise.
Note however that because of computational limitations, we could not use the same set of simulations in both cases. This led to different best fits when propagating uncertainties or not (nevertheless, the differences are well within the error bars).

It is clear that after the first iteration where the estimation of some parameters is somewhat biased (most notably $\Delta'_x$, $a_{c12}$ and $a_{c13}$), the iterative correction of instrumental parameters allows us to recover the input values, well within the required accuracy. In practice, based on these results, we iterate on instrumental parameters while estimates from one iteration to the next do not vary significantly, with a maximum number of iterations set to five \cite{hardycqg6}. Note also that the $a_{c12}$ and $a_{c13}$ parameters, though correctly estimated here, are not measured in real data \cite{hardycqg6, metriscqg9} since they are close to their required value from construction, and the noise (higher than in these simulations) does not allow for a precise measurement.

The discussions in Sect. \ref{ssect_errprop} and \ref{sect_appErrProp} show that error propagation is not trivial in our iterative calibration. The comparison of each instrumental parameter's lines in Table \ref{tab_errprop} (the first one corresponding to the case where we propagate errors, the second one to the case where we ignore error propagation) shows that we can ignore error propagation when estimating instrumental parameters. 

The last line of Table \ref{tab_errprop} gives the measurement of the E\"otv\"os parameter after correction of the estimated instrumental parameters. It is clear that we are able to accurately recover the input EPV signal with the expected precision. Just like in the case of instrumental parameters, the uncertainty on the E\"otv\"os parameter is not impacted by the uncertainty propagation from one iteration to the next. We can therefore reliably ignore the error propagation in our iterative estimations.

\section{Conclusion} \label{sect_conclusion}

In this paper, we summarised MICROSCOPE's data analysis process. Building on the measurement equation, we showed how we can either estimate, model or ignore instrumental parameters. The estimation of those that cannot be ignored nor modelled is performed through an iterative weighted least square fit in the frequency domain. We provided an extensive characterisation of {\it inpainting}, a gap-filling technique that we adapted to MICROSCOPE. This exercice, based on worst-case scenario numerical simulations, shows the behavior of the algorithm in a complex data-processing pipeline, and can be useful to other experimental data analyses.

Using well-controlled numerical simulations, we then showed that our iterative least-square method is robust to estimate instrumental parameters and correct for them in order to reliably measure the E\"otv\"os ratio. In particular, we showed that our main least-square estimator is intrinsically unbiased. Combined with our previous works \cite{baghi15,baghi16,berge15b,pires16}, those results prove that our data analysis pipeline allows us to reliably measure an EPV even in the presence of missing data and instrumental imperfections.

Finally, we discussed the non-trivial, but important, problem of uncertainties propagation in our pipeline. Like in any precision experiment, the problem of uncertainty estimation is central to the MICROSCOPE data analysis. We showed that not only does our iterative least-square technique provide correct error bars, but we can also safely ignore the problem of uncertainty propagation throughout iterations. This paper therefore justifies the results provided in Refs. \cite{touboul17,touboul19,touboulcqg0,metriscqg9}.

\ack
This work makes use of technical data from the CNES-ESA-ONERA-CNRS-OCA MICROSCOPE mission, and has received financial support from ONERA and CNES.
We thank Bruno Christophe, Bernard Foulon and Isabelle Petitbon, as well as the members of the MICROSCOPE Science Working Group for useful discussions. Special thanks go to Pierre Fayet and Pieter Visser for detailed comments on the manuscript. JB and SP acknowledge the financial support of the UnivEarthS Labex program at Sorbonne Paris Cit\'e (ANR-10-LABX-0023 and ANR-11-IDEX-0005-02).

\appendix
\section{In-flight instrumental parameters estimation} \label{app_calibration}

Table \ref{table_calib} lists the parameters that we can calibrate, and summarizes the techniques used to perform their estimation. The third column of the table lists the precision and accuracy required on the estimation of each parameter to satisfy the overall MICROSCOPE goal to reach the $10^{-15}$ level for the E\"otv\"os parameter; the fourth column shows the maximum value allowed by design of the instrument; those numbers were obtained with a performance analysis, whose description is beyond the scope of this paper. 
See Ref. \cite{hardycqg6} for a complete description of which parameters can be estimated in flight.

\begin{table}[t]
\caption{Parameters that can be calibrated in orbit, method to perform their calibration, and precision required to satisfy MICROSCOPE's objective. Oscillations are given along or about axes in the instrument frame. $f_{\rm EP}$ is the frequency of the test, where we may expect to detect a WEP violation. The fourth column gives the maximal allowed value by design. See Ref. \cite{hardy13b,hardycqg6} for details.}
\label{table_calib}
\begin{center}
\resizebox{\textwidth}{!}{%
\begin{tabular}{c|l|c|c}
Parameter & Calibration technique & Aimed precision & Max. allowed value \\
\hline
$\Delta'_{x}$ & No satellite maneuver; take advantage of Earth's GGT signal at $2 f_{\rm EP}$ & $0.1 \mu m$ & 20 $\mu$m\\
$\Delta'_{y}$ & Angular oscillation of the satellite about the $z$-axis & $2 \mu m$  & 20 $\mu$m\\
$\Delta'_{z}$ &  No satellite maneuver; take advantage of Earth's GGT signal at $2 f_{\rm EP}$  & $0.1 \mu m$  & 20 $\mu$m \\
$a_{c12}$ & Angular oscillation of the satellite about the $x$-axis and linear oscillation & $9\times{}10^{-4}$rad  & $2.6\times{}10^{-3}$rad \\
 & of the test mass along the $z$-axis & \\
$a_{c13}$ & Angular oscillation of the satellite about the $x$-axis and linear oscillation & $9\times{}10^{-4}$rad   & $2.6\times{}10^{-3}$rad\\
 & of the test mass along the $y$-axis & \\
$a'_{d11}$ & Linear oscillation of the satellite along the $x$-axis & $1.5\times{}10^{-4}$ & 0.01 \\
$a_{d12}$ & Linear oscillation of the satellite along the $y$-axis & $5\times{}10^{-5}$rad  & $1.5\times{}10^{-3}$rad \\
$a_{d13}$ & Linear oscillation of the satellite along the $z$-axis & $5\times{}10^{-5}$rad   & $1.5\times{}10^{-3}$rad\\
$K_{2xx}^{(d)} / \left(K_{1x}^{(c)}\right)^2$ & Linear oscillation of the satellite along the $x$-axis & 250 s$^2$/m & 14000 s$^2$/m \\
\hline
\end{tabular}}
\end{center}
\end{table}

\section{Error propagation through least-square estimation} \label{sect_appErrProp}

This appendix presents a pedagogical derivation of error propagation when correcting the measured differential acceleration from estimated instrumental parameters. We show two different approaches: the first one is a traditional variational approach, while the second one computes the full variance of the calibrated differential acceleration. We show that the latter provides a more complete view of the effects of propagating uncertainties when correcting for (estimated) instrumental parameter.

\subsection{Uncorrected differential acceleration and estimated parameters}

The uncorrected differential acceleration (Eq. \ref{eq_xacc}) can be re-written as:

\begin{multline} \label{eqdiff}
\Gamma^d(t) = 2\Gamma_\delta(t) + \sum_{j,k=1}^3 (T_{jk}(t) - In_{jk}(t)) a_{c1j} \Delta_k(t) \\
+ 2 \sum_{k=1}^3(-1)^k \left[a_{c1i}\dot\Delta_j(t) - a_{c1j}\dot\Delta_i(t) \right]_{i<j, i,j\neq k} \Omega_k \\
-\sum_{k=1}^3 a_{c1k} \ddot\Delta_k(t) + 2\sum_{k=1}^3a_{d1k} \tilde{\Gamma_k^c}(t) \\
+ K_{21} \left( \tilde{\Gamma}^{(1)}(t) \right)^2 - K_{22} \left( \tilde{\Gamma}^{(2)}(t) \right)^2 + 2 n_d(t),
\end{multline}

\noindent where $\Gamma_\delta(t)$ gathers terms depending on the E\"otv\"os ratio $\delta$, and where the indices $i,j,k$ are either in (1,2) or ($x$,$y$,$z$) depending on the variable they index. In Eq. (\ref{eqdiff}), we ignore angular-to-linear couplings.
In this equation, the common-mode and individual accelerations, marked by a tilde, are noise-corrected (e.g. $\tilde{\Gamma_k^c}(t) = \Gamma_k^c(t) - n_c(t)$).

The common-mode and differential-mode noises are assumed normally distributed with variance $\sigma_d^2 = \sigma_1^2 + \sigma_2^2$ and $\sigma_c^2 = (\sigma_1^2 + \sigma_2^2)/4$, where $\sigma_k^2$ is the variance for the $k$th ($k=1,2$) sensor's noise.

The parameters estimated before correction are the following: $a_{c1k}$, $a_{d1k}$, $\Delta_k$, $K_{2k}$.
In the following, we assume that their estimators $\hat{a}_{c1k}$, $\hat{a}_{d1k}$, $\hat\Delta_k$, $\hat{K}_{2k}$ are unbiased and of variance $\sigma_{ac1k}$, etc.

Note that although $a_{c11}$ cannot be estimated, it is possible to consider it as an unknown parameter, and assume it is a random variable of mean 1 and variance given by its specifications (or an educated guess). For the sake of clarity, we shall not pursue this possibility here.

\subsection{Synthetic uncorrected and calibrated accelerations}

In order to simplify the computation of the calibrated differential acceleration, of its expectation value and of its variance after propagating the errors on the estimation of instrumental parameters, we define a synthetic (uncorrected) differential acceleration as:

\begin{equation}
\Gamma^d(t) = \Gamma_\delta(t) + \kappa_a a Y(t) + \kappa_{bc} bc Z(t) + \kappa_k k \tilde{\Gamma^c}(t) + l_1 \left( \tilde{\Gamma}^{(1)} \right)^2 - l_2 \left( \tilde{\Gamma}^{(2)} \right)^2 + n_d(t),
\end{equation}
where $a$, $b$, $c$, $k$, $l_1$ and $l_2$ are estimated, $\kappa_i$ are numerical (constant) factors, $Y(t)$ and $Z(t)$ are deterministic signals that can be measured or modeled, and we kept the common-mode and individual accelerations unchanged. Although $b$ and $c$ may be degenerate, we will assume that they are not correlated.
This equation encompasses all types of terms found in Eq. (\ref{eqdiff}).

The calibrated synthetic acceleration can then be shown to be:
\begin{multline} \label{n2bsynth}
\Gamma_{\rm cal}^d(t) = \Gamma_\delta(t) + \kappa_a (a-\hat{a}) Y(t) + \kappa_{bc}(bc - \hat{b}\hat{c}) Z(t) + \kappa_k(k - \hat{k}) \tilde\Gamma^c(t) \\
+(l_1 - \hat{l_1}) \left[ \tilde\Gamma^{(1)}(t)\right]^2 - 2\hat{l_1} \tilde\Gamma^{(1)}(t)n_1(t) - (l_2 - \hat{l_2}) \left[ \tilde\Gamma^{(2)}(t)\right]^2 + 2\hat{l_2} \tilde\Gamma^{(2)}(t)n_2(t) \\
+\frac{1}{2} \left[ (n_1(t) - n_2(t)) - \kappa_k \hat{k} (n_1(t) + n_2(t))\right] - \hat{l_1}n_1^2(t) + \hat{l_2} n_2^2(t)
\end{multline}

In what follows, we assume that we wish to estimate the E\"otv\"os ratio $\delta$ (hence, the assumption that the estimates of the instrumental parameters are unbiased). The discussion can readily be generalized to the estimation of any instrumental parameter given priors on the others.

\subsection{Error propagation. Method 1: variational approach}

We first use a variational approach to compute the variance of the calibrated differential acceleration (see e.g. \cite{ku66}).

\subsection{Expectation value of calibrated differential acceleration}

Under the assumption that all estimates are unbiased, the expectation value of Eq. (\ref{n2bsynth}) is
\begin{equation}  \label{eq_ev1}
\mathrm{E}[\Gamma_{\rm cal}^d(t)] = \Gamma_\delta(t) - \hat{l_1} \sigma_1^2 + \hat{l_2} \sigma_2^2.
\end{equation}

It should be noted that the calibration entails a constant non-zero bias. However, the same bias is applied at all times $t$, and therefore does not affect the estimation of the amplitude of the WEP violation signal.


\subsubsection{Error propagation}

Taking the sum of partial derivatives of Eq. (\ref{n2bsynth}) and assuming that $|\Gamma^{(1)}| \ll{}1$ and $|\Gamma^{(2)}| \ll{}1$, we get:
\begin{multline} \label{eq_synth_variance}
\sigma_{\Gamma_{\rm cal}}^2 = \kappa_a^2 Y^2(t) \sigma_a^2 + \kappa_{bc} Z^2(t) \hat{b}^2 \hat{c}^2 \left( \frac{\sigma_b^2}{\hat{b}} + \frac{\sigma_c^2}{\hat{c}^2} \right) + \kappa_k^2 \left[ \Gamma^c(t)\right]^2 \sigma_k^2 \\
+ \left[ \frac{1-\kappa_k \hat{k}}{2} - 2 \hat{l_1} \Gamma^{(1)}(t) \right]^2 \sigma_1^2 + \left[2\hat{l_2} \Gamma^{(2)(t)} - \frac{1+\kappa_k \hat{k}}{2} \right]^2 \sigma_2^2
\end{multline}
Note that in Eq. (\ref{eq_synth_variance}), the common-mode and individual accelerations that appear are the effectively measured (hence, noisy) ones, contrary to those which appeared in the uncorrected acceleration.

\subsubsection{Actual calibrated acceleration: Expectation value and variance}

Returning to the actual MICROSCOPE accelerations, Eqs. (\ref{eq_ev1}) and (\ref{eq_synth_variance}) give:

\begin{equation} \label{eq_evalcal}
\mathrm{E}[\Gamma_{\rm cal}^d(t)] = \Gamma_\delta(t) - \frac{1}{2} (\hat{K}_{21} \sigma_1^2 + \hat{K}_{22} \sigma_2^2)
\end{equation}

\noindent and
\begin{multline} \label{eq_n2bvar}
\sigma_{\Gamma_{\rm cal}(t)}^2 = \frac{1}{4} \sum_{j,k=1}^3 \left[T_{jk}(t) - In_{jk}(t)\right]^2 \hat{a}_{c1j}^2 \hat\Delta_k^2 \left( \frac{\sigma_{\Delta_k}^2}{\hat{\Delta}_k^2} + \frac{\sigma_{ac1j}^2}{\hat{a_{c1j}}^2} \right) \\
+ \sum_{k=1}^3 \left[ \dot\Delta_j(t)^2 \sigma_{ac1i}^2 + \dot\Delta_i(t)^2 \sigma_{ac1j}^2 \right]_{i,j\neq k, i<j} \Omega_k(t)^2 \\
+ \frac{1}{4} \sum_{k=1}^3 \ddot\Delta_k(t)^2 \sigma_{ac1k}^2 
+ \sum_{k=1}^3 \left[ \Gamma^c(t)\right]^2 \sigma_{ad1k}^2 \\
+ \left[ \frac{1-\sum_{k=1}^3 \hat{a}_{d1k}}{2} - 2\hat{K}_{21} \Gamma^{(1)}(t) \right]^2 \sigma_1^2 \\
+ \left[ 2\hat{K}_{22} \Gamma^{(2)}(t) - \frac{1+\sum_{k=1}^3 \hat{a}_{d1k}}{2} \right]^2 \sigma_2^2.
\end{multline}

Note that the uncertainties on the quadratic factors $\sigma_{K2i}^2$ do not appear in this equation. This is because we assumed that $|\Gamma_{(i)}| \ll{}1$.

Eq. (\ref{eq_n2bvar}) shows that the instrumental parameters themselves and the uncertainties on their estimation bring up two different effects on the variance of the calibrated differential acceleration:
\begin{itemize}
\item extra-contributors to total variance: uncertainties on the estimated instrumental parameters add up to the noise, therefore increasing the calibrated differential acceleration's variance, and increasing the uncertainty on the E\"otv\"os parameter estimated with a least-square fit of the calibrated differential acceleration. Those extra contributions can be taken into account simply by adding them in the data covariance passed to the least-square fit. Another method (more computationally expensive) is to use Monte Carlo simulations where we vary the correction terms within their allowed bounds.
\item modification of the noise: the calibrated differential acceleration noise variance is not simply $\sigma_d^2 = \sigma_1^2 + \sigma_2^2$ as that of the uncorrected differential acceleration noise. Instead, it is modified by the presence of non-zero differential parameters ($a_{d1k}$) that couple to the common-mode acceleration and of non-zero quadratic factors that couple to the individual accelerations. The knowledge of the best estimates for those instrumental parameters allows us to quantify the modification of the noise (which, given the values of the involved parameters, remains negligible). However, Eq. (\ref{eq_n2bvar}) is valid only for a given set of instrumental parameters' estimators, and does not tell us anything about the distribution of the corrected differential acceleration noise, and hence on the uncertainty on the calibrated differential acceleration noise. A more general approach is necessary (see below).
\end{itemize}

\subsection{Error propagation. Method 2: general variance analysis of the calibrated differential acceleration}

For this analysis, we first go back to our synthetic model (\ref{n2bsynth}), whose variance we directly compute.

\subsubsection{Synthetic model}

Under the same assumptions as for Eq. (\ref{eq_synth_variance}), we find (with a straightforward but tedious algebra):

\begin{multline} \label{eq_synth_variance2}
\mathrm{Var}[\Gamma_{\rm cal}(t)] = \kappa_a^2 Y^2(t) \sigma_a^2 + \kappa_{bc}^2 Z^2(t) \hat{b}^2 \hat{c}^2 \left[\frac{\sigma_b^2 \sigma_c^2}{\hat{b}^2 \hat{c}^2} + \frac{\sigma_b^2}{\hat{b}^2} + \frac{\sigma_c^2}{\hat{c}^2} \right] + \kappa_k^2 \left[\tilde{\Gamma^c}(t) \right]^2 \sigma_k^2 + l_1^2 \sigma_1^4 + l_2 \sigma_2^4 \\
+ \frac{1}{4} \left[ 1+\kappa_k(\kappa_k \sigma_k^2 - 2k + k^2) + 16(\sigma_{l1}^2+l_1^2) \left[\tilde{\Gamma}^{(1)}(t)\right]^2 - 2(1-\kappa_kk)l_1 \tilde{\Gamma}^{(1)}(t)  \right] \sigma_1^2 \\
+ \frac{1}{4} \left[ 1+\kappa_k(\kappa_k \sigma_k^2 + 2k + k^2) + 16(\sigma_{l_2}^2+l_2^2) \left[\tilde{\Gamma}^{(2)}(t)\right]^2 - 2(1-\kappa_kk)l_2 \tilde{\Gamma}^{(2)}(t)  \right] \sigma_2^2.
\end{multline}

Note that although we assume $\sigma_{1,2} \ll{}1$, we do not ignore the $\sigma_{1,2}^4$ terms since the $l_{1,2}$ terms may be significant (they encode the quadratic factors).

We can observe that the contributors from the error propagation (those which do not contribute to the noise) are the same as in Eq. (\ref{eq_synth_variance}), except for the $\sigma_b^2 \sigma_c^2/\hat{b}^2 \hat{c}^2$ term, that pops-up here since we did not assume that $b$ and $c$ were independent. However, for MICROSCOPE, we always have $\sigma_{b,c}^2 \ll{}(b,c)^2$, so we will ignore this term in the following, so that the contribution to the uncertainty on the E\"otv\"os parameter from error propagation is the same in our two analyses (Eqs. (\ref{eq_synth_variance}) and (\ref{eq_synth_variance2})).

However, the noise term is much more complex in Eq. (\ref{eq_synth_variance2}) than in Eq. (\ref{eq_synth_variance}). This is because it encompasses the uncertainty on the estimated instrumental parameters. In that sense, whereas Eq. (\ref{eq_synth_variance}) tells us how the noise is modified when we correct the differential acceleration for a given set of instrumental parameters, Eq. (\ref{eq_synth_variance2}) tells us how the modified noise is distributed. It provides a more conservative uncertainty propagation, and should be favored when analyzing data.

\subsubsection{Actual calibrated differential acceleration}

Ignoring the  $\sigma_b^2 \sigma_c^2/\hat{b}^2 \hat{c}^2$ term from Eq. (\ref{eq_synth_variance2}), the total variance for the actual MICROSCOPE acceleration is
\begin{multline} \label{eq_variance2}
\mathrm{Var}[\Gamma_{\rm cal}(t)] = \frac{1}{4} \sum_{j,k=1}^3 \left[T_{jk}(t) - In_{jk}(t)\right]^2 \hat{a}_{c1j}^2 \hat\Delta_k^2 \left( \frac{\sigma_{\Delta_k}^2}{\hat{\Delta}_k^2} + \frac{\sigma_{ac1j}^2}{\hat{a}_{c1j}^2} \right) \\
+ \sum_{k=1}^3 \left[ \dot\Delta_j(t)^2 \sigma_{ac1i}^2 + \dot\Delta_i(t)^2 \sigma_{ac1j}^2 \right]_{i,j\neq k, i<j} \Omega_k(t)^2 \\
+ \frac{1}{4} \sum_{k=1}^3 \ddot\Delta_k(t)^2 \sigma_{ac1k}^2 
+ \sum_{k=1}^3 \left[ \Gamma^c(t)\right]^2 \sigma_{ad1k}^2 \\
+\frac{1}{4} \left\{ 1 + \sum_{k=1}^3 \sigma_{ad1k}^2 - 2\sum_{k=1}^3a_{d1k} +\sum_{k=1}^3 a_{d1k}^2 + 16 \left(\sigma_{K21}^2 + K_{21}^2 \right) \left[ \tilde{\Gamma}^{(1)} \right]^2 \right. \\
\left. - 2\left(1 - \sum_{k=1}^3 a_{d1k} \right) K_{21} \tilde{\Gamma}^{(1)}\right\} \sigma_1^2 \\
+\frac{1}{4} \left\{ 1 + \sum_{k=1}^3 \sigma_{ad1k}^2 + 2\sum_{k=1}^3a_{d1k} +\sum_{k=1}^3 a_{d1k}^2 + 16 \left(\sigma_{K22}^2 + K_{22}^2 \right) \left[ \tilde{\Gamma}^{(2)} \right]^2 \right. \\
\left. - 2\left(1 + \sum_{k=1}^3 a_{d1k} \right) K_{22} \tilde{\Gamma}^{(2)}\right\} \sigma_2^2 \\
+ K_{21}^2 \sigma_1^4 + K_{22}^2 \sigma_2^4
\end{multline}

The noise term in Eq. (\ref{eq_variance2}) makes it clear that the $a_{d1k}$ parameters and quadratic factors, as well as the uncertainty on their estimation, modify the measured noise.

Although in practice the uncertainty on those parameters can be seen as bringing an uncertainty on the noise (the noise is corrected by some uncertain estimates of the parameters), Eq. (\ref{eq_variance2}) combines those uncertainties to provide an upper bound of the measured noise. That is, it provides the most conservative variance of the calibrated differential acceleration. Therefore, it should be used to completely take into account the errors in the estimate of instrumental parameters, and should be included in the least-square fit.

Note that in this equation, the values of $a_{d1k}$ and $K_{2i}$ are not their estimates, but their ``real'' value. In practice, assuming that our estimates are unbiased, or at least give a correct order of magnitude of the real value, we can replace them by their estimated values.

\section*{References}
\bibliographystyle{iopart-num}
\bibliography{mic17}

\end{document}